\documentclass[a4paper,fleqn,usenatbib]{mnras}
\usepackage{newtxtext,newtxmath}
\usepackage[T1]{fontenc}
\usepackage{ae,aecompl}
\usepackage{graphicx}
\usepackage{float}
\usepackage{caption}
\usepackage{multicol}
\usepackage{multirow}
\usepackage{comment}

\title[Radio and X-ray variability of RQ Seyfert galaxies]{Quasi-simultaneous observations of radio and X-ray variability in three radio-quiet Seyfert galaxies}

\author[Sina Chen et al.]{
Sina Chen $^{1}$ \thanks{E-mail: sina.chen@campus.technion.ac.il},
Ari Laor $^{1}$,
and Ehud Behar $^{1}$.
\\
$^{1}$ Physics Department, Technion, Haifa 32000, Israel \\}

\date{Accepted XXX, Received YYY, in original form ZZZ.}

\pubyear{2022}

\begin{document}
\label{firstpage}
\pagerange{\pageref{firstpage}--\pageref{lastpage}}
\maketitle

\begin{abstract}

Radio variability in some radio-quiet (RQ) active galactic nuclei suggests emission from regions close to the central engine, possibly the outer accretion disc corona.
If the origins of the radio and the X-ray emission are physically related, their emission may be temporarily correlated, possibly with some time delays.
We present the results of quasi-simultaneous radio and X-ray monitoring of three RQ Seyfert galaxies, Mrk~110, Mrk~766, and NGC~4593, carried out with the Very Large Array at 8.5~GHz over a period of about 300 days, and with the Rossi X-ray Timing Explorer at 2--10~keV over a period of about 2000 days.
The radio core variability is likely detected in the highest resolution (A configuration) observations of Mrk~110 and NGC~4593, with a fractional variability amplitude of 6.3\% and 9.5\%, respectively.
A cross-correlation analysis suggests an apparently strong (Pearson $r = -0.89$) and highly significant correlation ($p = 1 \times 10^{-6}$) in Mrk~110, with the radio lagging the X-ray by 56 days.
However, a further analysis of the $r$ values distribution for physically unrelated long time delays, reveals that this correlation is not significant.
This occurs since the Pearson correlation assumes white noise, while both the X-ray and the radio light curves follow red noise, which dramatically increases the chance, by a factor of $\sim 10^3$, to get extremely high $r$ values in uncorrelated data sets.
A significantly longer radio monitoring with a higher sampling rate, preferably with a high-resolution fixed radio array, is required in order to reliably detect a delay.

\end{abstract}


\section{Introduction}

Active galactic nuclei (AGN) are believed to be powered by a super-massive black hole (SMBH).
They can be divided into radio-loud (RL) and radio-quiet (RQ) sub-classes.
The radio loudness is defined as the ratio of the 5~GHz radio flux density to the optical B-band flux density $R = S_{5\rm{GHz}} / S_{4400\text{\AA}}$ \citep{Kellermann1989}, where RL objects have $R > 10$.
The radio loudness is also defined using the ratio of the 5~GHz radio luminosity to the 2--10~keV X-ray luminosity $R = L_{5\rm{GHz}} / L_{2-10\rm{keV}}$ \citep{Terashima2003}, where RL objects have $\log R > -3.5$ \citep{Panessa2007,Laor2008}.
The radio emission in RL sources is typically $10^3$ times higher than that in RQ sources.
Such luminous radio emission is generally attributed to a relativistic jet via synchrotron radiation \citep{Begelman1984}.
However, only a fraction ($\sim$ 10\%) of AGN are RL.
The majority of the AGN population are RQ and are comparatively understudied in radio.
The nature of the radio emission in RQ AGN may be attributed to nuclear star formation, an AGN-driven wind, a low-power jet, accretion disc coronal emission, free-free emission, and possible a combination of these processes \citep{Panessa2019}.

Radio spectra and images are the key to distinguish among these physical scenarios.
Star formation generally produces steep spectra via optically thin synchrotron emission \citep{Condon1992,Kimball2011,Condon2013}.
The radio images show host galaxy scale diffuse emission surrounding the nucleus.
In contrast, jets and winds typically exhibit radio emission with a linear structure, which possibly extends from pc to kpc scales.
The spectrum of winds is expected to be steep, and the shocks that accelerate particles to relativistic energies likely occur on pc scale and above \citep{Jiang2010,Zakamska2016}.
The jet emission can be steep, if it is produced by extended emission on larger scales, but it can also be flat, if it mostly originates from a compact jet base \citep{Falcke1995,Jarvis2019}.
If the radio emission is of coronal origin, it is expected to be compact on sub-pc scale, with a flat or inverted optically thick spectrum \citep{Laor2008,Raginski2016}.
However, it may well be hard to distinguish among different emission mechanisms, in particular when the radio emission is unresolved.

Timing study is another important tool to characterize the physical origin, as it can provide very tight constraints on the size of the radio emitting region, and provide a physical connection to other wavebands.
Radio variability of RQ AGN, including quasars, Seyfert galaxies, and low-luminosity AGN, was indeed reported in the literature \citep{Falcke2001,Barvainis2005,Anderson2005,Mundell2009,Panessa2021}, such as variability at 8.5~GHz on time scales from a few days to a month.
Possible variability was also detected at higher frequencies $\sim$ 100~GHz \citep{Doi2011,Baldi2015,Behar2020}, which further supports the expected compact nature of the high frequency emission.

RQ AGN, in particular Seyfert galaxies, exhibit fast X-ray variability on time scales of hours, which indicates that the X-ray emission comes from a region close to the SMBH, probably the accretion disc corona \citep{Markowitz2004}.
The first evidence that the radio emission in RQ AGN is related to the X-ray emission comes from the correlation of $L_{\rm R} / L_{\rm X} \sim 10^{-5}$ where $L_{\rm R}$ is the radio luminosity at 5~GHz and $L_{\rm X}$ is the X-ray luminosity at 0.2--20~keV.
This radio and X-ray luminosity correlation holds for coronally active cool stars \citep{Guedel1993}, and as found by \citet{Laor2008}, also holds for the Palomar-Green (PG) quasars \citep{Boroson1992}, low luminosity AGN, and even ultra luminous X-ray sources, which suggests that the radio emission in RQ AGN may also be of coronal origin.
In RL AGN, $L_{\rm R} / L_{\rm X} \sim 10^{-2}$, and the radio emission has a jet origin.
Millimeter (mm) observations of RQ AGN found that the spectra at mm band display a compact and optically thick core superimposing on a steep power-law dominated at low frequencies \citep{Inoue2014,Behar2015,Doi2016a,Behar2018}.
This confirms that the corona indeed can produce synchrotron radiation with a flat radio spectrum up to mm band \citep{Raginski2016,Inoue2018}.

If the radio and X-ray emission are physically related, there will likely be a correlation between the radio and X-ray variability.
In coronally active cool stars, a radio flare often precedes a rise in the X-ray flux, which is called the Neupert effect \citep{Neupert1968,Gudel2002}.
The likely mechanism is a coronal magnetic reconnection event, which leads to electrons acceleration and produces the radio flare through synchrotron emission.
Most of the electron energy is dissipated by heating the corona, which leads to the X-ray emission.
During this process, only a small fraction of energy released is emitted in radio, and the majority is thermalized in the corona and emitted in X-ray, which leads to $L_{\rm R} / L_{\rm X} \sim 10^{-5}$.
Can we see the Neupert effect in RQ AGN?
Simultaneous mm and X-ray monitoring of Seyfert galaxies shows similar fractional variability in both bands \citep{Baldi2015,Behar2020}, which suggests that the mm and X-ray emission have the same physical origin, possibly the accretion disc corona.

In this paper we report the results of quasi-simultaneous radio and X-ray monitoring of three RQ Seyfert galaxies. The radio observations were carried out with the Very Large Array (VLA) in the X-band (8.5~GHz). The X-ray observations are part of a long-term ($\sim$ 6 year) monitoring carried out with the Rossi X-ray Timing Explorer (RXTE) at 2--10~keV. We present the targets in Section 2, the observations and data reduction in Section 3, and the results in Section 4. The discussion and summary are provided in Sections 5 and 6. Throughout this work, we adopt a standard $\Lambda$CDM cosmology with a Hubble constant $H_0 = 70$~km~s$^{-1}$~Mpc$^{-1}$, $\Omega_{\Lambda} = 0.73$, and $\Omega_{\rm M} = 0.27$ \citep{Komatsu2011}.

\section{The targets}

The radio monitoring targets are three nearby RQ Seyfert 1 galaxies, Mrk~110, Mrk~766, and NGC~4593, which were monitored in an ongoing long-term RXTE monitoring campaign \citep{Markowitz2004}.
Their X-ray variability amplitudes on one day time scale are about 10--15\% \citep{Markowitz2004,Boller2007}, and they are relatively bright in the radio (a few mJy), which facilitates the detection of radio variability.
The three objects were observed at 1.4, 5, and 8.5 GHz with the VLA A configuration in previous studies.
Their radio morphologies are generally point sources at 8.5~GHz, which enable us to detect the radio emission from the central cores.
The RXTE monitoring shows fast X-ray variability, which raise the possibility of detecting radio variability.
The redshift, black hole mass, Eddington ratio, flux densities at 1.4, 5, and 8.5 GHz with the VLA A configuration, spectral indices at 1.4--5~GHz and 5--8.5~GHz, and radio loudness are listed in Table~\ref{target}.
The Eddington ratios are computed via $L/L_{\rm Edd} = L_{\rm bol} / L_{\rm Edd}$ where $L_{\rm bol} = 9 L_{5100\text{\AA}}$ or $L_{\rm bol} = 10 L_{2-10\rm{keV}}$ and $L_{\rm Edd} = 1.3 \times 10^{38} (M_{\rm BH} / M_{\odot})$.
The spectral indices are calculated by modeling the spectrum with a power-law and using the definition of $\alpha = \frac{\log(S_2/S_1)}{\log(\nu_2/\nu_1)}$, where $S_1$ and $S_2$ are the flux densities at the observed frequencies $\nu_1$ and $\nu_2$, respectively.
The radio loudness $\log R = \log (L_{5\rm{GHz}} / L_{2-10\rm{keV}})$ is calculated using the 5~GHz radio flux in literature (Table~\ref{target}) and the 2--10~keV X-ray flux in the RXTE database (Table~\ref{X-ray}).
The $\log R$ in the three objects ranges from $-5.7$ to $-4.6$, which follows the $L_{\rm R}$ and $L_{\rm X}$ correlation in RQ AGN.

\begin{table*}
\centering
\caption{Previous studies of the targets.}
\label{target}
\begin{tabular}{cccccccccccccc}
\hline
\hline
\multirow{2}{*}{Target} & \multirow{2}{*}{$z$} & scale & \multirow{2}{*}{$M_{\rm BH}$} & \multirow{2}{*}{$L/L_{\rm Edd}$} & \multirow{2}{*}{Ref.} & $S_{1.4}$ & $S_5$ & $S_{8.5}$ & \multirow{2}{*}{Ref.} & \multirow{2}{*}{$\alpha_{1.4-5}$} & \multirow{2}{*}{$\alpha_{5-8.5}$} & \multirow{2}{*}{$\log \frac{L_{5\rm{GHz}}}{L_{2-10\rm{keV}}}$} & \multirow{2}{*}{$\log L_{8.5}$} \\
& & (kpc arcsec$^{-1}$) & & & & (mJy) & (mJy) & (mJy) & & & & & \\
(1) & (2) & (3) & (4) & (5) & (6) & (7) & (8) & (9) & (10) & (11) & (12) & (13) & (14) \\
\hline
Mrk~110 & 0.0353 & 0.753 & 6.9 & 0.47 & a,b & 5.8 & 1.9 & 1.7 & f & $-$0.88 & $-$0.21 & $-5.5$ & 28.71 \\
Mrk~766 & 0.0129 & 0.271 & 6.2 & 0.42 & c & 39.3 & 14.8 & 8.7 & g,h,i & $-$0.77 & $-$1.00 & $-4.6$ & 28.59 \\
NGC~4593 & 0.0090 & 0.188 & 6.7 & 0.19 & d,e & 2.1 & 1.6 & 2.0 & j,k,l & $-$0.21 & 0.42 & $-5.7$ & 27.62 \\
\hline
\end{tabular}
\flushleft{\textbf{Notes.}
Columns: (1) name, (2) redshift, (3) physical scale, (4) logarithm of black hole mass in units of $M_{\odot}$, (5) Eddington ratio, (6) references for black hole mass and Eddington ratio, (7) flux density at 1.4~GHz observed with VLA A configuration, (8) flux density at 5~GHz observed with VLA A configuration, (9) flux density at 8.5~GHz observed with VLA A configuration, (10) references for the flux densities, (11) spectral index at 1.4--5~GHz, (12) spectral index at 5--8.5~GHz, (13) radio loudness using the ratio of the 5~GHz radio flux to the 2--10~keV X-ray flux, (14) logarithm of luminosity density at observed 8.5~GHz in units of erg~s$^{-1}$~Hz$^{-1}$.
References:
(a). \citet{Kaspi2000},
(b). \citet{Davis2011},
(c). \citet{Bentz2009},
(d). \citet{Williams2018},
(e). \citet{Peterson2004a},
(f). \citet{Kukula1998},
(g). \citet{Kukula1995},
(h). \citet{Nagar1999},
(i). \citet{Berton2018},
(j). \citet{Ulvestad1984},
(k). \citet{Thean2000},
(l). \citet{Schmitt2001}.}
\end{table*}

\textbf{Mrk~110:} The spectral indices indicate that the radio emission is optically thin ($\alpha < -0.5$) at 1--5~GHz and is optically thick ($\alpha > -0.5$) at 5--9~GHz (see Table~\ref{target}). The radio images at 1.4 and 5 GHz display extended emission which may be associated with an outflow \citep{Kukula1998,Berton2018}. However, such an outflow is not detected at 8.5~GHz probably because this structure has a steep spectrum and/or is resolved out at higher resolution. Thus, only a compact core is detected at 8.5~GHz. This target was additionally observed with the Very Long Baseline Array (VLBA) at 1.7~GHz having a flux density of $S_{1.7} = 1.1$~mJy and a compact core of $< 10$~pc with a brightness temperature of $T_{\rm B} > 6.5 \times 10^7$~K which excludes a thermal origin \citep{Doi2013}. In addition, a VLBA monitoring at 5~GHz reveals significant variability on a time scale of a single day \citep{Panessa2021}.

\textbf{Mrk~766:} The spectral indices suggest that the radio emission is dominated by an optically thin source at 1--9~GHz (see Table~\ref{target}). The radio images at 1.4, 5, and 8.5 GHz show slightly extended emission which may be related to an outflow. Additionally, this target harbours a strong water maser emission at 22~GHz which seems to favour the outflow interpretation \citep{Tarchi2011}. Observations with the VLBA on mas resolution at 1.7~GHz give a flux density of $S_{1.7} = 2.5$~mJy. The radio core has a slightly extended morphology with a brightness temperature of $T_{\rm B} = 8.4 \times 10^7$~K, which again indicates a synchrotron origin \citep{Doi2013}. However, it was not detected with the Very Long Baseline Interferometry (VLBI) at 22~GHz, with a flux upper limit of 7~mJy \citep{Doi2016b}.

\textbf{NGC~4593:} The radio emission at 1--9~GHz is dominated by an optically thick source as indicated by the spectral indices (see Table~\ref{target}). Radio images show a point source at sub-arcsec and arcsec resolutions. But low-resolution observations with the VLA D configuration reveal the presence of extended radio emission, possibly lobes \citep{Gallimore2006}, on the host galaxy scale (30$^{\prime\prime}$, $\sim$ 5~kpc), which may be resolved out at higher resolutions.

\section{Observation and data reduction}

\subsection{Radio with the VLA}

The VLA monitoring projects were carried out at 8.5~GHz in three consecutive semesters using the A, B, and D configurations (see Table~\ref{obs}).
There are a total of 59 useful epochs, including 55 for Mrk~110, 52 for Mrk~766, and 55 for NGC~4593.
The integrated time is about 10--14 min for Mrk~110, 3--5 min for Mrk~766, and 10--16 min for NGC~4593 in each epoch.

\begin{table}
\centering
\caption{Observational details of the VLA monitoring projects at 8.5~GHz.}
\label{obs}
\begin{tabular}{cccc}
\hline
\hline
\multirow{2}{*}{Proposal code} & Date & \multirow{2}{*}{Configuration} & Resolution \\
& (DD/MM/YYYY) & & (arcsec) \\
\hline
VLA/08B-182 & 27/06/2008 -- 13/09/2008 & VLA-D & 7.2 \\
VLA/08C-123 & 26/09/2008 -- 12/01/2009 & VLA-A & 0.2 \\
VLA/09A-149 & 20/02/2009 -- 19/05/2009 & VLA-B & 0.6 \\
\hline
\end{tabular}
\end{table}

We reduced the data using the Common Astronomy Software Applications (CASA) version 5.5.0. A standard flux density calibrator 3C 286 was used for all targets. We started from setting a flux density scale using the task SETJY and the observation of the flux calibrator. The gain solutions were determined using the task GAINCAL and the gain calibrators of each target. We then bootstrapped the flux density scale onto the complex gain calibrators using the task FLUXSCALE and transferred the solutions to the sources using the task APPLYCAL. A baseline-based correction was also performed on the flux calibrator using the task BLCAL. We then applied both the antenna-based and baseline-based calibration solutions to the data using the task APPLYCAL again. Finally, the calibrated source data was split from the multi-source measurement set using the task SPLIT. This flux calibration process was performed on each observation epoch.

To produce the radio maps of each source, we used a cell size of 0.05$^{\prime\prime}$ for A and B configurations and 0.5$^{\prime\prime}$ for D configuration to properly sample the beam at different configurations. The maps were created in a region of 2048 $\times$ 2048 pixels for A and B configurations and 1024 $\times$ 1024 pixels for D configuration centered on the target coordinates to check for the presence of nearby sources. We modeled the main target along with the nearby sources (if present) using the CLEAN algorithm in all spectral windows to avoid the contamination from sidelobes. We used a natural weighting to create the images. No self-calibration was applied as the sources are not sufficiently bright.

The sources were selected in the central region with a flux level higher than 3$\sigma$ and excluding the extended structure if present.
We modeled the sources with a Gaussian fit on the image plane to obtain the integrated and peak flux densities, $S_{\rm int}$ and $S_{\rm p}$, centered at 8.5~GHz.
The errors of each measurement $\sigma_{\rm err}$ were returned by CASA.
The background noises $\sigma_{\rm bg}$ were estimated in a source-free region.
These parameters are listed in Table~\ref{flux}.
There are several observations without reliable measurements.
The reasons mainly are (a) data flagged due to RFI, (b) only a few antennas available, (c) bad weather, (d) no data on flux or phase calibrators.

\begin{table*}
\centering
\scriptsize
\caption{Integrated and peak flux densities, and background noise of Mrk~110, Mrk~766, and NGC~4593 in the VLA monitoring observations at 8.5~GHz.}
\label{flux}
\begin{tabular}{cccccccccccc}
\hline
\hline
\multirow{3}{*}{Date} & \multirow{2}{*}{MJD} & \multirow{3}{*}{Configuration} & \multicolumn{3}{c}{Mrk~110} & \multicolumn{3}{c}{Mrk~766} & \multicolumn{3}{c}{NGC~4593} \\
& & & $S_{\rm int} \pm \sigma_{\rm err}$ & $S_{\rm p} \pm \sigma_{\rm err}$ & $\sigma_{\rm bg}$ & $S_{\rm int} \pm \sigma_{\rm err}$ & $S_{\rm p} \pm \sigma_{\rm err}$ & $\sigma_{\rm bg}$ & $S_{\rm int} \pm \sigma_{\rm err}$ & $S_{\rm p} \pm \sigma_{\rm err}$ & $\sigma_{\rm bg}$ \\
& (day) & & (mJy) & (mJy~beam$^{-1}$) & ($\mu$Jy~beam$^{-1}$) & (mJy) & (mJy~beam$^{-1}$) & ($\mu$Jy~beam$^{-1}$) & (mJy) & (mJy~beam$^{-1}$) & ($\mu$Jy~beam$^{-1}$) \\
\hline
2008-06-27 & 54644 & \multirow{20}{*}{VLA-D} & 3.77 $\pm$ 0.17 & 2.84 $\pm$ 0.08 & 42 & 10.3 $\pm$ 0.16 & 10.12 $\pm$ 0.09 & 75 & 2.32 $\pm$ 0.13 & 1.61 $\pm$ 0.06 & 48 \\
2008-07-02 & 54649 & & 4.02 $\pm$ 0.13 & 2.95 $\pm$ 0.06 & 42 & 10.19 $\pm$ 0.1 & 10.09 $\pm$ 0.05 & 75 & 2.31 $\pm$ 0.12 & 1.72 $\pm$ 0.06 & 45 \\
2008-07-05 & 54652 & & 3.94 $\pm$ 0.16 & 2.99 $\pm$ 0.07 & 52 & 10.99 $\pm$ 0.41 & 10.1 $\pm$ 0.22 & 86 & 2.33 $\pm$ 0.17 & 1.67 $\pm$ 0.07 & 66 \\
2008-07-09 & 54656 & & 3.75 $\pm$ 0.13 & 2.82 $\pm$ 0.06 & 42 & 10.26 $\pm$ 0.16 & 9.89 $\pm$ 0.09 & 61 & 2.41 $\pm$ 0.13 & 1.69 $\pm$ 0.06 & 42 \\
2008-07-11 & 54658 & & 3.89 $\pm$ 0.14 & 2.97 $\pm$ 0.07 & 45 & 10.4 $\pm$ 0.14 & 10.04 $\pm$ 0.08 & 73 & 2.39 $\pm$ 0.11 & 1.63 $\pm$ 0.05 & 45 \\
2008-07-30 & 54677 & & 3.93 $\pm$ 0.14 & 3.07 $\pm$ 0.07 & 58 & 10.19 $\pm$ 0.34 & 10.02 $\pm$ 0.19 & 90 & 2.43 $\pm$ 0.09 & 1.73 $\pm$ 0.04 & 44 \\
2008-08-01 & 54679 & & 4.03 $\pm$ 0.16 & 3.1 $\pm$ 0.07 & 49 & 10.5 $\pm$ 0.09 & 10.27 $\pm$ 0.05 & 80 & 2.54 $\pm$ 0.1 & 1.76 $\pm$ 0.04 & 39 \\
2008-08-05 & 54683 & & 3.84 $\pm$ 0.15 & 3.02 $\pm$ 0.07 & 47 & 10.17 $\pm$ 0.2 & 9.91 $\pm$ 0.11 & 72 & 2.66 $\pm$ 0.15 & 1.94 $\pm$ 0.07 & 43 \\
2008-08-08 & 54686 & & 3.95 $\pm$ 0.11 & 3.07 $\pm$ 0.05 & 52 & 10.38 $\pm$ 0.21 & 10.28 $\pm$ 0.12 & 95 & 2.63 $\pm$ 0.11 & 1.84 $\pm$ 0.05 & 38 \\
2008-08-15 & 54693 & & 4.01 $\pm$ 0.11 & 3.09 $\pm$ 0.05 & 45 & 10.19 $\pm$ 0.24 & 10.07 $\pm$ 0.14 & 87 & 2.65 $\pm$ 0.1 & 1.86 $\pm$ 0.05 & 40 \\
2008-08-19 & 54697 & & 3.98 $\pm$ 0.14 & 3.04 $\pm$ 0.06 & 48 & 10.32 $\pm$ 0.17 & 10.1 $\pm$ 0.09 & 82 & 2.33 $\pm$ 0.15 & 1.8 $\pm$ 0.07 & 43 \\
2008-08-21 & 54699 & & 3.85 $\pm$ 0.2 & 2.78 $\pm$ 0.09 & 48 & 10.24 $\pm$ 0.13 & 9.89 $\pm$ 0.07 & 75 & 2.48 $\pm$ 0.15 & 1.84 $\pm$ 0.07 & 38 \\
2008-08-26 & 54704 & & 3.98 $\pm$ 0.16 & 2.9 $\pm$ 0.07 & 49 & 10.44 $\pm$ 0.27 & 10.21 $\pm$ 0.15 & 89 & 2.52 $\pm$ 0.13 & 1.84 $\pm$ 0.06 & 36 \\
2008-08-28 & 54706 & & - & - & - & - & - & - & - & - & - \\
2008-08-29 & 54707 & & - & - & - & 10.33 $\pm$ 0.25 & 10.13 $\pm$ 0.14 & 97 & 2.52 $\pm$ 0.14 & 1.82 $\pm$ 0.07 & 41 \\
2008-09-02 & 54711 & & 3.72 $\pm$ 0.21 & 2.85 $\pm$ 0.1 & 77 & 10.38 $\pm$ 0.5 & 9.88 $\pm$ 0.27 & 115 & 2.4 $\pm$ 0.19 & 1.84 $\pm$ 0.09 & 76 \\
2008-09-07 & 54716 & & 3.8 $\pm$ 0.16 & 2.96 $\pm$ 0.07 & 83 & 10.52 $\pm$ 0.47 & 9.89 $\pm$ 0.26 & 82 & 2.5 $\pm$ 0.24 & 1.77 $\pm$ 0.11 & 51 \\
2008-09-12 & 54721 & & 3.98 $\pm$ 0.24 & 2.98 $\pm$ 0.11 & 57 & 10.32 $\pm$ 0.24 & 9.8 $\pm$ 0.13 & 93 & 2.64 $\pm$ 0.21 & 1.84 $\pm$ 0.09 & 50 \\
2008-09-13 & 54722 & & 3.8 $\pm$ 0.13 & 2.92 $\pm$ 0.06 & 45 & 10.8 $\pm$ 0.14 & 10.2 $\pm$ 0.08 & 72 & 2.63 $\pm$ 0.15 & 1.9 $\pm$ 0.07 & 47 \\
2008-09-26 & 54735 & & - & - & - & - & - & - & - & - & - \\
\hline
2008-11-10 & 54780 & \multirow{19}{*}{VLA-A} & 2.18 $\pm$ 0.14 & 1.98 $\pm$ 0.08 & 69 & 8.87 $\pm$ 0.3 & 5.29 $\pm$ 0.12 & 81 & 1.03 $\pm$ 0.21 & 0.84 $\pm$ 0.1 & 57 \\
2008-11-16 & 54786 & & - & - & - & - & - & - & - & - & - \\
2008-11-21 & 54791 & & 2.22 $\pm$ 0.11 & 1.95 $\pm$ 0.06 & 49 & - & - & - & 1.33 $\pm$ 0.09 & 1.1 $\pm$ 0.04 & 55 \\
2008-12-01 & 54801 & & 2.21 $\pm$ 0.07 & 2.01 $\pm$ 0.04 & 46 & 9.66 $\pm$ 0.72 & 4.93 $\pm$ 0.26 & 268 & 1.3 $\pm$ 0.11 & 1.21 $\pm$ 0.06 & 50 \\
2008-12-05 & 54805 & & 2.12 $\pm$ 0.08 & 2.02 $\pm$ 0.04 & 44 & 8.55 $\pm$ 0.61 & 4.35 $\pm$ 0.22 & 234 & 1.45 $\pm$ 0.05 & 1.32 $\pm$ 0.03 & 48 \\
2008-12-09 & 54809 & & 2.08 $\pm$ 0.11 & 1.96 $\pm$ 0.06 & 61 & 9.36 $\pm$ 0.94 & 4.24 $\pm$ 0.31 & 276 & 1.38 $\pm$ 0.09 & 1.29 $\pm$ 0.05 & 66 \\
2008-12-14 & 54814 & & 2.05 $\pm$ 0.06 & 1.96 $\pm$ 0.03 & 59 & - & - & - & 1.46 $\pm$ 0.07 & 1.45 $\pm$ 0.04 & 83 \\
2008-12-17 & 54817 & & 2.0 $\pm$ 0.08 & 1.78 $\pm$ 0.04 & 36 & 8.92 $\pm$ 0.41 & 5.0 $\pm$ 0.16 & 230 & 1.25 $\pm$ 0.06 & 1.25 $\pm$ 0.04 & 37 \\
2008-12-18 & 54818 & & 2.07 $\pm$ 0.09 & 1.76 $\pm$ 0.05 & 52 & - & - & - & 1.32 $\pm$ 0.05 & 1.32 $\pm$ 0.05 & 50 \\
2008-12-20 & 54820 & & 2.0 $\pm$ 0.11 & 1.7 $\pm$ 0.06 & 47 & 8.3 $\pm$ 1.2 & 5.81 $\pm$ 0.53 & 702 & 1.36 $\pm$ 0.06 & 1.25 $\pm$ 0.03 & 47 \\
2008-12-24 & 54824 & & 1.95 $\pm$ 0.1 & 1.73 $\pm$ 0.06 & 52 & 8.43 $\pm$ 0.55 & 5.27 $\pm$ 0.23 & 262 & 1.39 $\pm$ 0.07 & 1.28 $\pm$ 0.04 & 43 \\
2008-12-27 & 54827 & & 2.04 $\pm$ 0.09 & 1.76 $\pm$ 0.05 & 66 & 8.91 $\pm$ 0.32 & 5.08 $\pm$ 0.13 & 84 & 1.34 $\pm$ 0.06 & 1.25 $\pm$ 0.03 & 60 \\
2009-01-02 & 54833 & & 2.02 $\pm$ 0.09 & 1.75 $\pm$ 0.05 & 45 & 8.43 $\pm$ 0.7 & 5.16 $\pm$ 0.29 & 293 & 1.58 $\pm$ 0.07 & 1.4 $\pm$ 0.04 & 45 \\
2009-01-03 & 54834 & & 1.9 $\pm$ 0.09 & 1.79 $\pm$ 0.05 & 55 & 8.75 $\pm$ 0.31 & 5.49 $\pm$ 0.13 & 81 & 1.32 $\pm$ 0.12 & 1.2 $\pm$ 0.06 & 53 \\
2009-01-04 & 54835 & & 1.85 $\pm$ 0.06 & 1.75 $\pm$ 0.03 & 53 & 8.52 $\pm$ 0.27 & 5.16 $\pm$ 0.11 & 89 & 1.33 $\pm$ 0.12 & 1.08 $\pm$ 0.06 & 50 \\
2009-01-07 & 54838 & & 1.97 $\pm$ 0.1 & 1.69 $\pm$ 0.05 & 56 & 8.54 $\pm$ 0.57 & 5.17 $\pm$ 0.23 & 289 & 1.39 $\pm$ 0.08 & 1.31 $\pm$ 0.04 & 47 \\
2009-01-09 & 54840 & & 1.82 $\pm$ 0.1 & 1.55 $\pm$ 0.05 & 56 & 9.01 $\pm$ 0.32 & 5.36 $\pm$ 0.13 & 75 & 1.67 $\pm$ 0.09 & 1.39 $\pm$ 0.04 & 48 \\
2009-01-10 & 54841 & & 1.88 $\pm$ 0.09 & 1.69 $\pm$ 0.04 & 46 & 9.1 $\pm$ 0.32 & 5.48 $\pm$ 0.13 & 75 & 1.73 $\pm$ 0.12 & 1.35 $\pm$ 0.06 & 47 \\
2009-01-12 & 54843 & & 1.81 $\pm$ 0.07 & 1.66 $\pm$ 0.04 & 47 & 8.93 $\pm$ 0.45 & 5.82 $\pm$ 0.19 & 245 & 1.7 $\pm$ 0.1 & 1.41 $\pm$ 0.05 & 54 \\
\hline
2009-02-20 & 54882 & \multirow{20}{*}{VLA-B} & 2.05 $\pm$ 0.06 & 1.84 $\pm$ 0.03 & 39 & 8.69 $\pm$ 0.16 & 7.78 $\pm$ 0.09 & 101 & 1.57 $\pm$ 0.04 & 1.51 $\pm$ 0.02 & 42 \\
2009-02-27 & 54889 & & 2.13 $\pm$ 0.06 & 1.91 $\pm$ 0.03 & 40 & 9.61 $\pm$ 0.17 & 8.41 $\pm$ 0.09 & 68 & 1.62 $\pm$ 0.05 & 1.55 $\pm$ 0.03 & 43 \\
2009-03-12 & 54902 & & - & - & - & 9.88 $\pm$ 0.37 & 8.22 $\pm$ 0.19 & 113 & - & - & - \\
2009-03-17 & 54907 & & 1.95 $\pm$ 0.13 & 1.9 $\pm$ 0.07 & 114 & - & - & - & 1.36 $\pm$ 0.07 & 1.36 $\pm$ 0.07 & 84 \\
2009-03-23 & 54913 & & 1.9 $\pm$ 0.1 & 1.73 $\pm$ 0.05 & 61 & 9.34 $\pm$ 0.33 & 8.06 $\pm$ 0.17 & 136 & 1.35 $\pm$ 0.08 & 1.33 $\pm$ 0.04 & 68 \\
2009-04-02 & 54923 & & 1.98 $\pm$ 0.07 & 1.78 $\pm$ 0.04 & 45 & - & - & - & 1.41 $\pm$ 0.12 & 1.33 $\pm$ 0.07 & 54 \\
2009-04-18 & 54939 & & 2.27 $\pm$ 0.13 & 1.81 $\pm$ 0.06 & 40 & 9.57 $\pm$ 0.44 & 8.13 $\pm$ 0.23 & 80 & 1.54 $\pm$ 0.07 & 1.44 $\pm$ 0.04 & 49 \\
2009-04-19 & 54940 & & 2.11 $\pm$ 0.09 & 1.84 $\pm$ 0.05 & 45 & 9.46 $\pm$ 0.21 & 8.16 $\pm$ 0.11 & 116 & 1.54 $\pm$ 0.15 & 1.26 $\pm$ 0.07 & 62 \\
2009-04-23 & 54944 & & 2.14 $\pm$ 0.07 & 1.86 $\pm$ 0.04 & 55 & 9.8 $\pm$ 0.2 & 8.44 $\pm$ 0.11 & 78 & 1.6 $\pm$ 0.08 & 1.53 $\pm$ 0.04 & 46 \\
2009-04-26 & 54947 & & 2.12 $\pm$ 0.1 & 1.74 $\pm$ 0.05 & 48 & 9.3 $\pm$ 0.33 & 8.11 $\pm$ 0.17 & 95 & - & - & - \\
2009-04-29 & 54950 & & 2.1 $\pm$ 0.05 & 1.88 $\pm$ 0.03 & 46 & 9.55 $\pm$ 0.18 & 8.2 $\pm$ 0.1 & 87 & 1.51 $\pm$ 0.11 & 1.48 $\pm$ 0.06 & 46 \\
2009-04-30 & 54951 & & 2.2 $\pm$ 0.06 & 1.91 $\pm$ 0.03 & 42 & 9.84 $\pm$ 0.24 & 8.29 $\pm$ 0.12 & 94 & 1.62 $\pm$ 0.06 & 1.51 $\pm$ 0.03 & 45 \\
2009-05-03 & 54954 & & 2.26 $\pm$ 0.32 & 1.83 $\pm$ 0.16 & 50 & 9.34 $\pm$ 0.53 & 7.71 $\pm$ 0.27 & 98 & 1.58 $\pm$ 0.19 & 1.23 $\pm$ 0.09 & 50 \\
2009-05-08 & 54959 & & 2.05 $\pm$ 0.06 & 1.89 $\pm$ 0.03 & 43 & 9.64 $\pm$ 0.13 & 8.43 $\pm$ 0.07 & 79 & 1.87 $\pm$ 0.15 & 1.44 $\pm$ 0.07 & 43 \\
2009-05-09 & 54960 & & 2.02 $\pm$ 0.06 & 1.91 $\pm$ 0.03 & 42 & 9.69 $\pm$ 0.23 & 8.23 $\pm$ 0.12 & 90 & 1.76 $\pm$ 0.1 & 1.46 $\pm$ 0.05 & 48 \\
2009-05-14 & 54965 & & 2.06 $\pm$ 0.06 & 1.88 $\pm$ 0.03 & 42 & 9.74 $\pm$ 0.14 & 8.52 $\pm$ 0.07 & 103 & 1.43 $\pm$ 0.03 & 1.43 $\pm$ 0.03 & 44 \\
2009-05-15 & 54966 & & 2.02 $\pm$ 0.08 & 1.86 $\pm$ 0.04 & 41 & 8.74 $\pm$ 0.28 & 7.5 $\pm$ 0.14 & 110 & 1.65 $\pm$ 0.09 & 1.42 $\pm$ 0.05 & 44 \\
2009-05-16 & 54967 & & 2.16 $\pm$ 0.14 & 1.85 $\pm$ 0.07 & 40 & 9.86 $\pm$ 0.37 & 8.33 $\pm$ 0.19 & 70 & 1.47 $\pm$ 0.08 & 1.26 $\pm$ 0.04 & 32 \\
2009-05-17 & 54968 & & 2.07 $\pm$ 0.16 & 1.85 $\pm$ 0.09 & 46 & 9.78 $\pm$ 0.23 & 8.47 $\pm$ 0.12 & 75 & 1.89 $\pm$ 0.14 & 1.37 $\pm$ 0.06 & 45 \\
2009-05-19 & 54970 & & 2.01 $\pm$ 0.07 & 1.81 $\pm$ 0.04 & 42 & 9.85 $\pm$ 0.25 & 8.43 $\pm$ 0.13 & 152 & 1.35 $\pm$ 0.03 & 1.35 $\pm$ 0.03 & 49 \\
\hline
\end{tabular}
\end{table*}

We note that in nine observations of Mrk~766, the integrated time is less than one minute. These are \textit{2008-12-01}, \textit{2008-12-05}, \textit{2008-12-09}, \textit{2008-12-17}, \textit{2008-12-20}, \textit{2008-12-24}, \textit{2009-01-02}, \textit{2009-01-07}, and \textit{2009-01-12}. The background noise in all these epochs is larger than 200 $\mu$Jy~beam$^{-1}$. We keep these measurements in the following analysis because their flux densities are in a reasonable range. However, they have to be taken with caution. In addition, in a few observations of NGC~4593, the integrated flux density is less than the peak flux density, either because the emission is completely unresolved with a given beam size, or there is an error in the estimate of the integrated flux density. These are \textit{2008-12-18}, \textit{2009-03-17}, \textit{2009-05-14}, and \textit{2009-05-19}. In these cases, we treated $S_{\rm p}$ as an upper limit for $S_{\rm int}$.

There are two additional objects, 1148$+$594 and 1244$+$408, included in the observations, which can be used as a check-source to verify if there are systematic errors introduced in the flux calibration.
We calibrated these two check-sources using the same procedure mentioned above, and measured their flux densities only in the A-configuration observations, since the variability is mainly detected in the A configuration (see Section 4.2).
The object 1148$+$594 has an average peak flux density of $416 \pm 9$~mJy with a standard deviation (STD) of 21~mJy.
This results in a variability of $\sim$ 5\%, which is consistent with the expected VLA flux calibration uncertainty \footnote{\url{https://science.nrao.edu/facilities/vla/docs/manuals/oss/performance/fdscale}} \citep{Perley2017}.
We thus consider that a source is variable if the amplitude of variability is larger than $\sim$ 5\%.
The other object 1244$+$408 was only detected in about half of the epochs observed with the A configuration, which hinders a detailed analysis.

\begin{figure*}
\centering
\includegraphics[width=\textwidth]{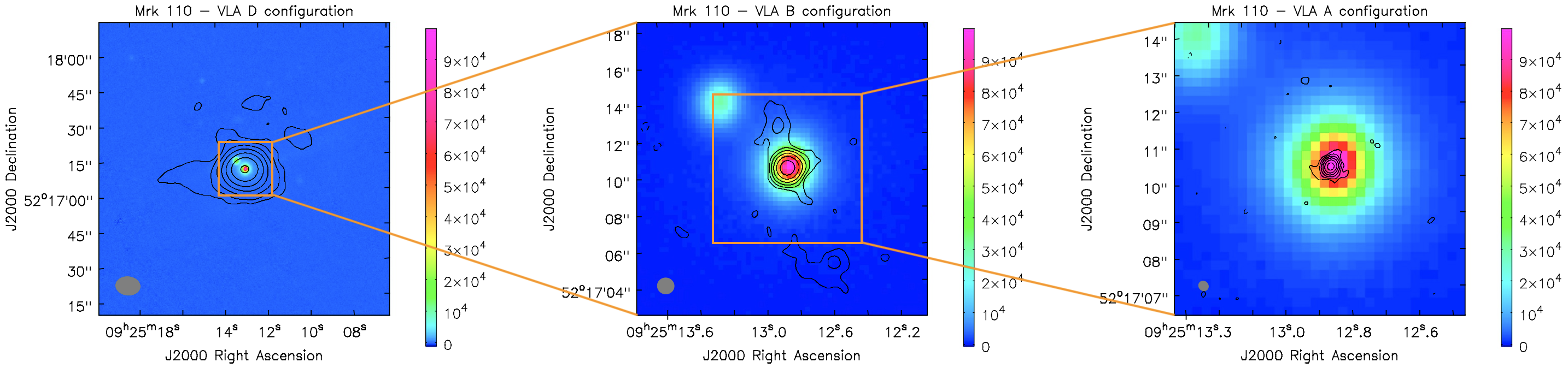}
\includegraphics[width=\textwidth]{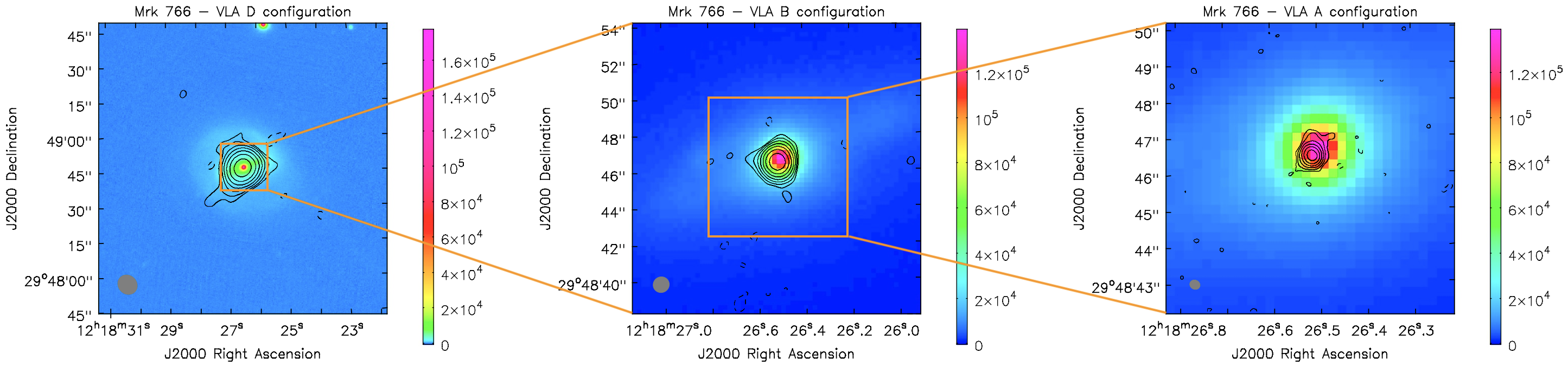}
\includegraphics[width=\textwidth]{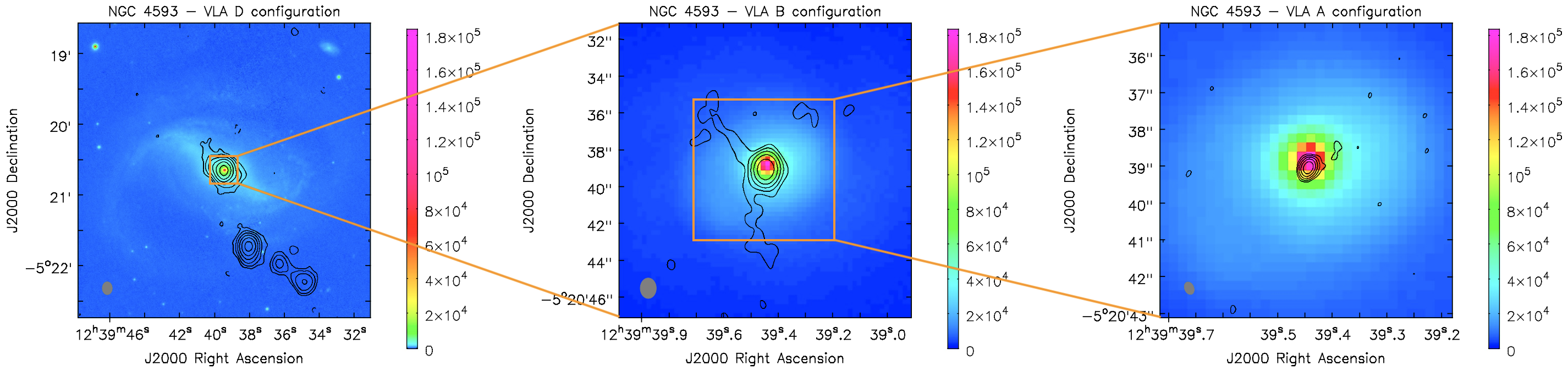}
\caption{The optical images (colour) overlaid with the radio contours of Mrk~110 (top row), Mrk~766 (middle row), and NGC~4593 (bottom row).
The optical images are a stack of {\em g}, {\em i}, {\em y} fields from the Pan-STARRS DR2, and the colour bars are in unit of counts.
The radio contours at 8.5~GHz are [$-$3, 3, 6, 12, 24, 48, 96, 192] $\times \sigma_{\rm bg}$ with the VLA D (left column), B (middle column), and A (right column) configurations in increasing resolutions. The beam is noted in the lower-left corner of each panel. In all objects the radio emission is dominated by an unresolved compact source.}
\label{overlap}
\end{figure*}

In addition, we combined all the observations in each configuration to obtain higher S/N radio images. The integrated and peak flux densities, and background noise of the combined images are listed in Table~\ref{allimages}.
We further overlaid the radio contours with the D, B, and A configurations in increasing resolutions on their optical images, as plotted in Figure~\ref{overlap}, in order to have an overview of the radio emission detected at different scales.
The optical images are from the Panoramic Survey Telescope and Rapid Response System (Pan-STARRS) Data Release 2 (DR2) \footnote{\url{https://outerspace.stsci.edu/display/PANSTARRS/}} \citep{Chambers2016,Flewelling2020}, and are a stack of {\em g}, {\em i}, {\em y} fields.

\begin{table}
\centering
\scriptsize
\caption{Integrated and peak flux densities, and background noise of the VLA combined images at 8.5~GHz.}
\label{allimages}
\begin{tabular}{ccccc}
\hline
\hline
\multirow{2}{*}{Target} & \multirow{2}{*}{Configuration} & $S_{\rm int} \pm \sigma_{\rm err}$ & $S_{\rm p} \pm \sigma_{\rm err}$ & $\sigma_{\rm bg}$ \\
& & (mJy) & (mJy~beam$^{-1}$) & ($\mu$Jy~beam$^{-1}$) \\
\hline
\multirow{3}{*}{Mrk~110} & VLA-A & 2.069 $\pm$ 0.051 & 1.842 $\pm$ 0.027 & 13.98 \\
& VLA-B & 2.109 $\pm$ 0.045 & 1.849 $\pm$ 0.024 & 11.61 \\
& VLA-D & 3.82 $\pm$ 0.10 & 2.784 $\pm$ 0.048 & 13.85 \\
\hline
\multirow{3}{*}{Mrk~766} & VLA-A & 8.75 $\pm$ 0.27 & 5.04 $\pm$ 0.10 & 34.32 \\
& VLA-B & 9.656 $\pm$ 0.096 & 8.248 $\pm$ 0.050 & 23.39 \\
& VLA-D & 10.35 $\pm$ 0.15 & 9.882 $\pm$ 0.082 & 29.54 \\
\hline
\multirow{3}{*}{NGC~4593} & VLA-A & 1.391 $\pm$ 0.025 & 1.260 $\pm$ 0.013 & 13.59 \\
& VLA-B & 1.581 $\pm$ 0.048 & 1.401 $\pm$ 0.026 & 15.04 \\
& VLA-D & 2.49 $\pm$ 0.11 & 1.696 $\pm$ 0.048 & 13.42 \\
\hline
\end{tabular}
\end{table}

\subsection{X-ray with the RXTE}

The X-ray light curves at 2--10~keV are from the RXTE AGN Timing $\&$ Spectral Database \footnote{\url{https://cass.ucsd.edu/~rxteagn/}}, which provides the community systematically-analyzed light curves and spectra for all AGN observed with the RXTE \citep{Rivers2013}.
Mrk~110 was monitored from 2005 to 2012 with a total of $\sim$ 1400 epochs, and Mrk~766 and NGC~4593 were both monitored from 2004 to 2012 with a total of $\sim$ 700 and $\sim$ 1300 epochs respectively.
The time steps were a couple of days for Mrk~110 and NGC~4593, and 4--5 days for Mrk~766.
The X-ray spectral model, the photon index $\Gamma$ where $F(E) \propto E^{-\Gamma}$, and the average 2--10~keV flux from the RXTE database are listed in Table~\ref{X-ray}.

\begin{table*}
\centering
\caption{The X-ray spectral model, the photon index $\Gamma$, and the average 2--10~keV flux from the RXTE database.}
\label{X-ray}
\begin{tabular}{cccc}
\hline
\hline
\multirow{2}{*}{Name} & \multirow{2}{*}{Model} & \multirow{2}{*}{$\Gamma$} & $F_{2-10\rm{keV}}$ \\
& & & (10$^{-11}$~erg~cm$^{-2}$~s$^{-1}$) \\
\hline
Mrk~110 & recorn $\ast$ phabs (powerlaw $+$ pexrav $+$ zgauss) & 1.80 $\pm$ 0.04 & 3.1 \\
Mrk~766 & recorn $\ast$ phabs (powerlaw $+$ pexrav $+$ zgauss) $\ast$ mtable\{warm\_abs\} & 2.33 $\pm$ 0.09 & 2.8 \\
NGC~4593 & recorn $\ast$ constant $\ast$ phabs (powerlaw $+$ pexrav $+$ zgauss) $\ast$ mtable\{warm\_abs\} & 1.85 $\pm$ 0.03 & 3.8 \\
\hline
\end{tabular}
\end{table*}

\section{Results}

\subsection{Images}

Figure~\ref{overlap} clearly shows that the radio emission at 8.5~GHz with the A configuration is mainly from the central core. The B-configuration images may include some extended emission, and the D-configuration images may show host galaxy scale emission and even an extra source.

\textbf{Mrk~110} exhibits a compact core in both A and B configurations, as their integrated flux densities agree within 1$\sigma$, which is consistent with previous studies (see Section 2). Faint extended linear structure is present in the B configuration, possibly associated with an outflow. The D-configuration image shows some extended emission, with an integrated flux density which is about twice the higher resolution images (see Table~\ref{allimages}).

\textbf{Mrk~766} displays only slightly extended emission in all configurations, as found in earlier studies (see Section 2). The integrated flux density in the A configuration takes up 91\% of that in the B configuration and 85\% of that in the D configuration.

\textbf{NGC~4593} shows a compact core in the A configuration, as found in earlier studies (see Section 2). The radio emission is relatively compact as the integrated flux density in the A configuration accounts for 88\% of that in the B configuration and 56\% of that in the D configuration. Extended emission is present in the B configuration, which is resolved out at higher resolution. The D-configuration image reveals slightly extended emission surrounding the central core, and three linear components to the southwest of the target. These three components have the angular separations of 68$^{\prime\prime}$, 93$^{\prime\prime}$, and 117$^{\prime\prime}$ from the central core, at $z$ = 0.0090, corresponding to the projected linear distances of 13.8~kpc, 18.9~kpc, and 23.8~kpc, respectively. They may be part of a one-sided deflected jet, or potentially an unrelated background source.

It is clear that the unresolved flux is significant or dominant in all configurations.
If the unresolved flux is indeed highly compact, it may show some radio variability.

\subsection{Light curves}

\begin{figure*}
\centering
\includegraphics[scale=0.7]{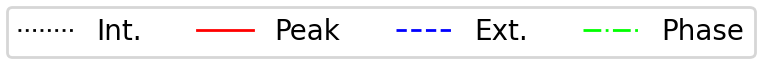}
\includegraphics[width=\textwidth, trim={2cm 0cm 3cm 0cm}, clip]{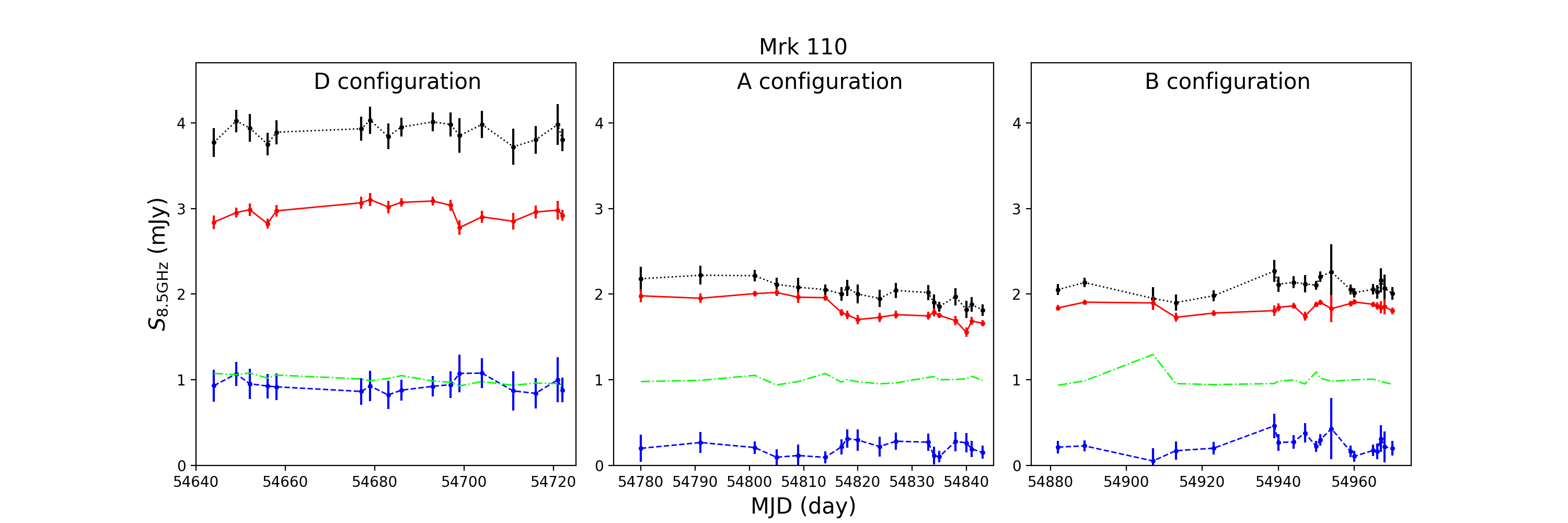}
\includegraphics[width=\textwidth, trim={2cm 0cm 3cm 0cm}, clip]{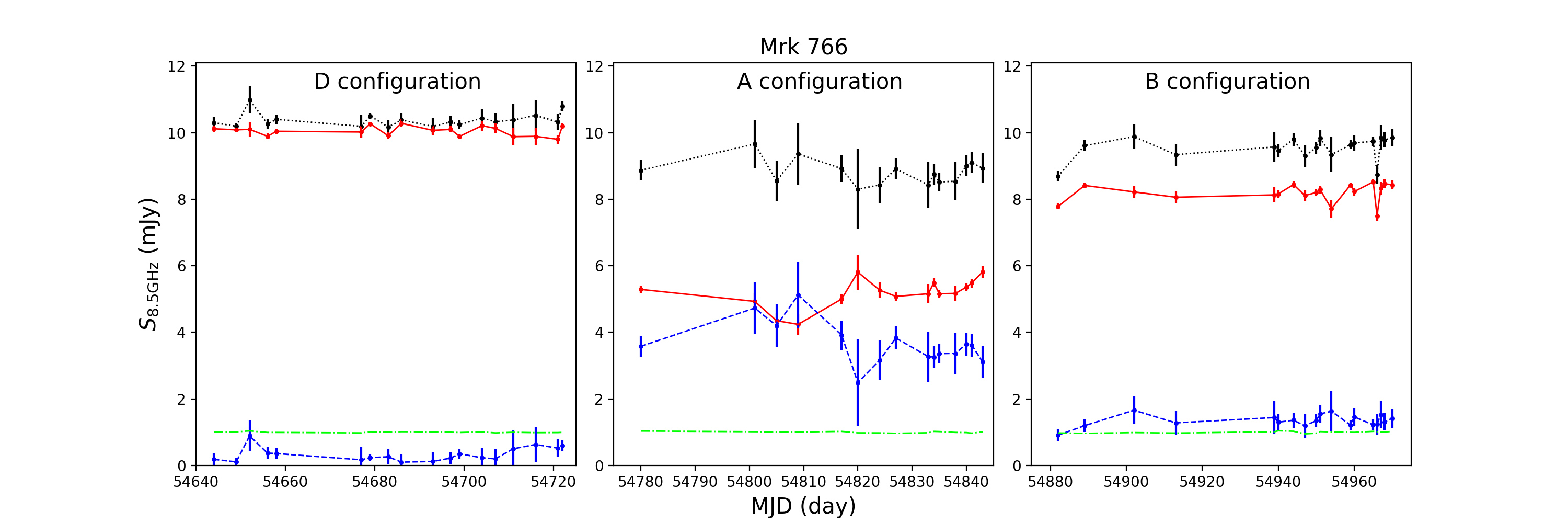}
\includegraphics[width=\textwidth, trim={2cm 0cm 3cm 0cm}, clip]{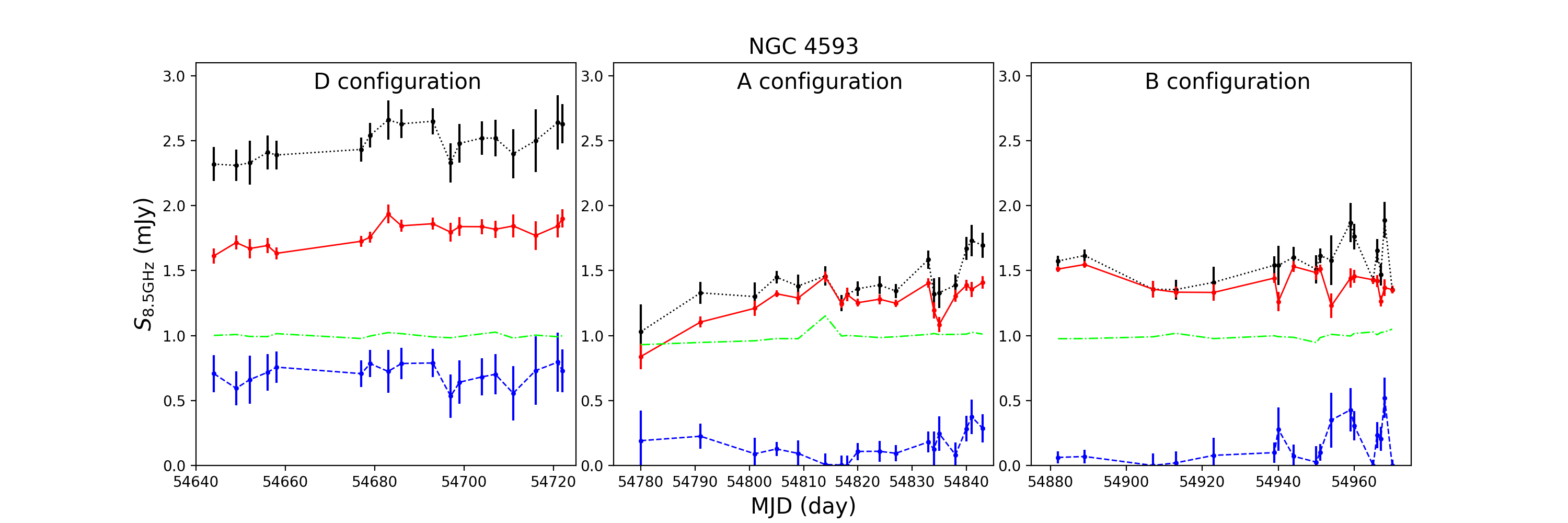}
\caption{The radio light curves of Mrk~110 (top row), Mrk~766 (middle row), and NGC~4593 (bottom row) at 8.5~GHz. The integrated, peak, and extended flux densities are shown in black dotted line, red solid line, and blue dashed line, respectively.
The lime dash-dot line represents the flux-to-mean ratio of the phase calibrator.
From left to right columns, the observations were carried out time-sequentially with D, A, and B configurations.
The extended emission cannot be variable, and its light curve provides an indication for the level of systematic flux errors.
The most significant variability is the peak flux density observed with the A configuration for Mrk~110 and NGC~4593, and the B configuration for Mrk~766.}
\label{LC_radio}
\end{figure*}

\begin{figure}
\centering
\includegraphics[width=.5\textwidth]{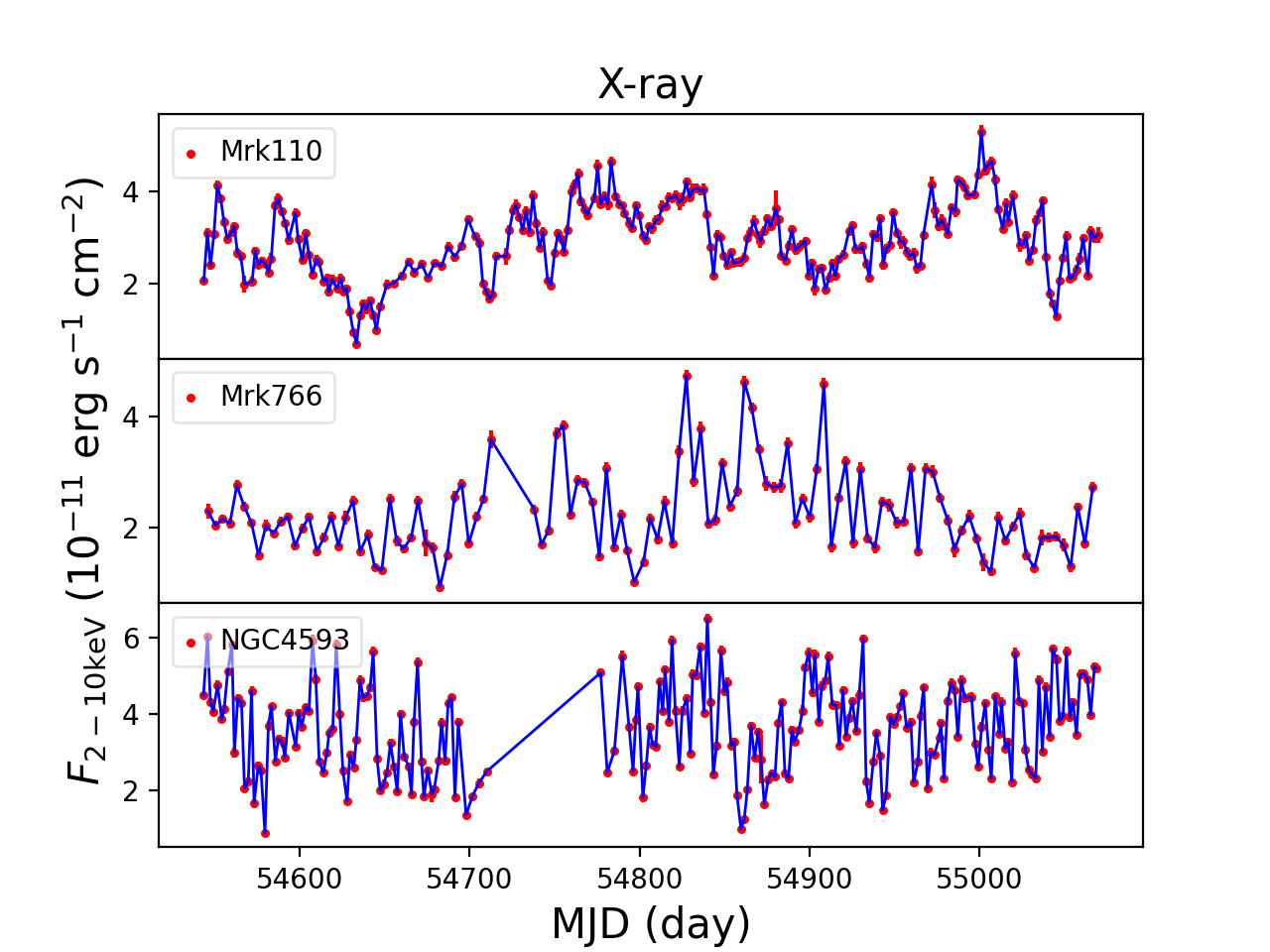}
\caption{The X-ray light curves of Mrk~110 (top panel), Mrk~766 (middle panel), and NGC~4593 (bottom panel) at 2--10~keV.
The lines between the data points are plotted just to guide the eyes.
Note that the variability pattern in both Mrk~766 and NGC~4593 looks like "white noise", i.e.\ no dependence of the variability amplitude on the time scales.
In Mrk~110 the variability amplitude increases with the variability time scales, characteristic of "red noise" variability.
The errors in the X-ray light curves are as small as the symbol size.}
\label{LC_X-ray}
\end{figure}

Figure~\ref{LC_radio} presents the radio light curves of Mrk~110, Mrk~766, and NGC~4593 with the D, A, and B configurations in time sequence. For each object in each configuration, the figure shows the integrated, peak, and extended flux densities. The extended flux density $S_{\rm ext}$ was calculated via $S_{\rm int} - S_{\rm p}$. In cases where the upper limit of $S_{\rm int}$ equals $S_{\rm p}$, we get $S_{\rm ext} = 0$, which means that no significant extended emission is detected.
The flux-to-mean ratios of the phase calibrators for each target and each observation are also plotted in the figure for comparison, in order to clarify whether the variability is genuine.
Since we are also interested in simultaneous radio and X-ray monitoring, the X-ray light curves are selected in a period of 100 days before the first radio observation ($t_1 - 100$) and 100 days after the last radio observation ($t_2 + 100$), which are plotted in Figure~\ref{LC_X-ray}.

The fractional variability amplitude $F_{\rm var}$ and its uncertainty $\sigma_{\rm var}$ of each light curve are calculated via \citep{Edelson2002,Vaughan2003,Markowitz2004}
\begin{equation}
F_{\rm var} = \frac{\sqrt{\displaystyle\sum_{i=1}^N \left[ \left( F_i - \left\langle F \right\rangle \right) ^2 - \sigma_i^2 \right] / N}}{\left\langle F \right\rangle}
\end{equation}
and
\begin{equation}
\sigma_{\rm var} = \sqrt{\left(\sqrt{\frac{\left \langle \sigma^2 \right \rangle}{N}} \cdot \frac{1}{\left \langle F \right \rangle}\right)^2 + \left(\sqrt{\frac{1}{2N}} \cdot \frac{\left \langle \sigma^2 \right \rangle}{\left \langle F \right \rangle ^2 F_{\rm var}}\right)^2}
\end{equation}
where $F_i$ and $\sigma_i = \sqrt{\sigma_{\rm err}^2 + \sigma_{\rm bg}^2}$ are the flux density and the uncertainty in the $i$-th epoch, $\left\langle F \right\rangle$ and $\left\langle \sigma \right\rangle$ are the mean flux density and its uncertainty, and $N$ is the number of data points. When $\sigma_i$ exceeds $F_i - \left\langle F \right\rangle$, the current epoch does not vary. When $\displaystyle\sum_{i=1}^N \left[ \left( F_i - \left\langle F \right\rangle \right) ^2 - \sigma_i^2 \right] < 0$, $F_{\rm var}$ is not well defined and the light curve does not vary.

We also estimated $\chi^2$ with respect to $\left\langle F \right\rangle$ in each light curve via \citep{Paolillo2004,Young2012,Lanzuisi2014}
\begin{equation}
\chi^2 = \sum_{i=1}^{N} \frac{\left( F_i - \left\langle F \right\rangle \right) ^2}{\sigma_i^2}.
\end{equation}
The reduced-$\chi^2$, $\chi^2_{\rm r}$ = $\chi^2$ / d.o.f, where d.o.f = $N-1$ is the number of the degrees of freedom, and the associated probability $p(\chi^2) = 1 - p(> \chi^2)$, give an idea of the significance of the variability in each light curve.
These parameters of the radio light curves in each configuration, and of the X-ray light curves, are listed in Table~\ref{variation}.

\begin{table*}
\centering
\caption{The amplitudes of variability in the radio at 8.5~GHz and the X-ray at 2--10~keV light curves.}
\label{variation}
\begin{tabular}{cccccccc}
\hline
\hline
Target & Configuration & $N$ & Comp. & $\left\langle F \right\rangle \pm \left\langle \sigma \right\rangle$ & $F_{\rm var} \pm \sigma_{\rm var}$ (\%) & $\chi^2_{\rm r}$ & $p(\chi^2)$ \\
(1) & (2) & (3) & (4) & (5) & (6) & (7) & (8) \\
\hline
\multirow{10}{*}{Mrk~110} & \multirow{3}{*}{VLA-A} & \multirow{3}{*}{18} & Int. & 2.01 $\pm$ 0.09 & 3.0 $\pm$ 1.2 & 1.70 & 0.04 \\
& & & Peak & 1.81 $\pm$ 0.05 & 6.3 $\pm$ 0.9 & 3.87 & 1.14 $\times$ 10$^{-7}$ \\
& & & Ext. & 0.21 $\pm$ 0.10 & - & 0.49 & 0.96 \\
& \multirow{3}{*}{VLA-B} & \multirow{3}{*}{19} & Int. & 2.09 $\pm$ 0.10 & - & 0.77 & 0.74 \\
& & & Peak & 1.85 $\pm$ 0.05 & - & 0.69 & 0.83 \\
& & & Ext. & 0.24 $\pm$ 0.11 & - & 0.64 & 0.87 \\
& \multirow{3}{*}{VLA-D} & \multirow{3}{*}{17} & Int. & 3.90 $\pm$ 0.16 & - & 0.41 & 0.98 \\
& & & Peak & 2.96 $\pm$ 0.07 & 1.0 $\pm$ 0.7 & 1.33 & 0.17 \\
& & & Ext. & 0.94 $\pm$ 0.17 & - & 0.20 & 1.00 \\
& X-ray & 239 & - & 2.94 $\pm$ 0.10 & 26.1 $\pm$ 0.2 & 65.48 & $< 10^{-15}$ \\
\hline
\multirow{10}{*}{Mrk~766} & \multirow{3}{*}{VLA-A} & \multirow{3}{*}{15} & Int. & 8.82 $\pm$ 0.53 & - & 0.36 & 0.98 \\
& & & Peak & 5.17 $\pm$ 0.21 & 4.5 $\pm$ 1.5 & 2.02 & 0.01 \\
& & & Ext. & 3.64 $\pm$ 0.57 & - & 0.69 & 0.78 \\
& \multirow{3}{*}{VLA-B} & \multirow{3}{*}{18} & Int. & 9.54 $\pm$ 0.27 & 1.7 $\pm$ 0.7 & 2.17 & 3.46 $\times$ 10$^{-3}$ \\
& & & Peak & 8.19 $\pm$ 0.14 & 2.5 $\pm$ 0.5 & 3.16 & 1.11 $\times$ 10$^{-5}$ \\
& & & Ext. & 1.35 $\pm$ 0.30 & - & 0.48 & 0.96 \\
& \multirow{3}{*}{VLA-D} & \multirow{3}{*}{18} & Int. & 10.38 $\pm$ 0.23 & - & 0.96 & 0.50 \\
& & & Peak & 10.05 $\pm$ 0.13 & - & 1.16 & 0.29 \\
& & & Ext. & 0.34 $\pm$ 0.27 & - & 0.64 & 0.86 \\
& X-ray & 118 & - & 2.27 $\pm$ 0.1 & 32.3 $\pm$ 0.4 & 60.45 & $< 10^{-15}$ \\
\hline
\multirow{10}{*}{NGC~4593} & \multirow{3}{*}{VLA-A} & \multirow{3}{*}{18} & Int. & 1.41 $\pm$ 0.09 & 9.0 $\pm$ 1.8 & 2.20 & 2.99 $\times$ 10$^{-3}$ \\
& & & Peak & 1.26 $\pm$ 0.05 & 9.5 $\pm$ 1.3 & 2.79 & 1.03 $\times$ 10$^{-4}$ \\
& & & Ext. & 0.15 $\pm$ 0.10 & - & 0.90 & 0.57 \\
& \multirow{3}{*}{VLA-B} & \multirow{3}{*}{18} & Int. & 1.56 $\pm$ 0.09 & 6.9 $\pm$ 1.6 & 2.59 & 3.33 $\times$ 10$^{-4}$ \\
& & & Peak & 1.40 $\pm$ 0.05 & 4.2 $\pm$ 1.2 & 2.25 & 2.21 $\times$ 10$^{-3}$ \\
& & & Ext. & 0.16 $\pm$ 0.10 & 58.7 $\pm$ 17.4 & 1.98 & 9.36 $\times$ 10$^{-3}$ \\
& \multirow{3}{*}{VLA-D} & \multirow{3}{*}{18} & Int. & 2.48 $\pm$ 0.14 & - & 0.78 & 0.72 \\
& & & Peak & 1.78 $\pm$ 0.06 & 1.9 $\pm$ 1.1 & 1.47 & 0.09 \\
& & & Ext. & 0.70 $\pm$ 0.16 & - & 0.24 & 1.00 \\
& X-ray & 221 & - & 3.63 $\pm$ 0.11 & 31.9 $\pm$ 0.2 & 136.72 & $< 10^{-15}$ \\
\hline
\end{tabular}
\flushleft{\textbf{Notes.} Columns: (1) name, (2) observation, (3) the number of data points, (4) the radio component, (5) the mean flux density and its uncertainty, (6) the fractional variability amplitude and its uncertainty, (7) the reduced-$\chi^2$ with respect to the mean flux density, (8) the probability of non-variability with the associated $\chi^2$.
The radio flux density at 8.5~GHz is in units of mJy. The X-ray flux at 2--10~keV is in units of 10$^{-11}$~erg~s$^{-1}$~cm$^{-2}$.}
\end{table*}

\textbf{Mrk~110:}
Significant variability is detected during the A-configuration observations.
The $S_{\rm p}$ variability amplitude is 6.3\% with a high significance level of $p = 10^{-7}$.
The variability pattern of $S_{\rm p}$ in the A configuration looks like a step function, with a change in flux density from $\sim$ 2~mJy before MJD 54815 to $\sim$ 1.7~mJy after MJD 54820. That is a flux drop by $\sim$ 15\% within 5 days.
No variability above the flux calibration uncertainty level (5\%) is detected in the B and D configurations.
Further, the $S_{\rm ext}$ light curves in all configurations do not show significant variability, as expected for extended emission, which indicates no significant variable systematic flux error.

\textbf{Mrk~766:}
No significant variability is detected in all observations.
Although the $S_{\rm p}$ variability in the B configuration is statistically significant ($p = 10^{-5}$), the amplitude is well below the flux calibration uncertainty level (5\%).
In the A configuration, the light curve of $S_{\rm p}$ appears to be anti-correlated with that of $S_{\rm ext}$, and consequently $S_{\rm int}$ shows no hint of variability.
The variability of $S_{\rm p}$ is likely an artefact of the point $+$ extended emission decomposition.
The non-detection of variability in the D configuration is not a low-resolution dilution effect, as $S_{\rm int}$ increases only slightly from the A to the D configurations.

\textbf{NGC~4593:}
Relatively large amplitude (9.5\%) and significant ($p = 10^{-4}$) variability is detected in $S_{\rm p}$ during the A-configuration observations.
Lower amplitude but still significant variability is detected in the B configuration (6.9\% and $p = 3 \times 10^{-4}$) for $S_{\rm int}$.
However, $S_{\rm ext}$ in the B configuration also shows significant and large amplitude variability, which is not physically possible.
Thus, the variability may reflect some systematic effects, possibly changes in the seeing which affects the spatial resolution and the detectable extended flux.
No significant variability is detected during the D-configuration observations.

In Mrk~110 and NGC~4593, we note that significant variability was not detected in the D-configuration observations, which could be a resolution effect.
The unresolved flux, $S_{\rm p}$, in the D configuration is a factor of 1.5--2 larger than that in the A and B configurations, which dilutes the potential variability of the compact radio source.

The X-ray variability amplitudes at 2--10~keV are 26.1\% for Mrk~110, 32.3\% for Mrk~766, and 31.9\% for NGC~4593 on time scales of days, and are significantly higher than their radio variability at 8.5~GHz.
Such rapid and strong X-ray variability is typical for Seyfert galaxies \citep{Edelson1996,Nandra1997,Edelson2002,Markowitz2004}.

\section{Discussion}

\subsection{Variability}

Potentially significant variability is detected in $S_{\rm p}$ of both Mrk~110 and NGC~4593 during the A-configuration observations, at levels of 6.3\% and 9.5\%, respectively.
The quoted flux calibration uncertainty is 5--10\% in the VLA X band, while our measurements of the check-source give a scatter of 5\% (see Section 3.1).
Thus the variability in Mrk~110 is only marginally significant, and in NGC~4593 it is more likely to be real.
No significant variability is detected in Mrk~766.

A recent study of Mrk~110 by \citet{Panessa2021} shows that the VLBA scale radio emission at 5~GHz varies significantly on time scales from days to weeks, and a simultaneous light curve observed with \textit{Swift} X-ray Telescope (XRT) also indicates strong X-ray variability.
It is interesting to note that the maximal VLBA flux change is $\sim$ 0.3~mJy (from 1 to 0.7 mJy) at 5~GHz.
This maximal flux change is the same as the maximal flux change measured here (from 2 to 1.7 mJy) with the VLA A configuration at 8.5~GHz.
The similar absolute flux changes at the VLBA 5~GHz and the VLA 8.5~GHz imply that the VLA spectral slope of $\alpha_{5-8.5} = -0.21$ likely applies to the VLBA core as well.
In addition, the overall VLBA variability amplitude of $F_{\rm var}$ = 12\% suggests an absolute flux variability which is similar to that in the VLA, as the mean flux of the VLA is about twice that of the VLBA, and its variability is about half ($F_{\rm var}$ = 6.3\%).
Specifically, about 1~mJy of the VLA flux may be produced on scales larger than the VLBA beam size $\sim$ 1--3~pc, and will not be variable.
Thus, although the VLA flux change of 6.3\% is only marginally significant, the similarity in amplitude to the VLBA flux changes suggests this variability is real.

Variability allows us to estimate the size of the optically thick emitting region based on its light crossing time.
In both Mrk~110 and NGC~4593, the synchrotron emission at 8.5~GHz is optically thick ($\alpha_{5-8.5} > -0.5$), and the size of the emitting region, $R$, can be derived from the observed radio and optical luminosity (assuming an equipartition magnetic field).
For example, following equation 22 in \citet{Laor2008}, we have $R_{\rm pc} = 0.47 L_{30}^{0.4} L_{46}^{0.1} \nu_{\rm GHz}^{-1}$ in pc, where $L_{30}$ is the radio luminosity density in units of $10^{30}$~erg~s$^{-1}$~Hz$^{-1}$, $L_{46}$ is the bolometric luminosity in units of $10^{46}$~erg~s$^{-1}$, and $\nu_{\rm GHz}$ is the observed frequency in GHz.
Using the luminosities in Table~\ref{target}, this expression gives $R$ = 14.9 light days in Mrk~110 and $R$ = 4.8 light days in NGC~4593.
The steepest variability in Mrk~110 is a decrease of 15\% (from 2 to 1.7 mJy) within 6 days, which is consistent with the scale of the light crossing time of $\sim$ 15 days.
Similarly, in NGC~4593 the steepest event is a drop of 20\% (from 1.4 to 1.1 mJy) within 2 days, again consistent with the scale of the light crossing time of $\sim$ 5 days.

In contrast, in Mrk~766 the synchrotron emission is optically thin ($\alpha_{5-8.5} = -1$).
The size of the 8.5~GHz emitting region must be well above the size if it is optically thick, which is $R$ = 11.2 light days.
The source remains optically thin down to 1.4~GHz, and if the overall flux at 1.4--8.5~GHz is produced by a single source, then the size of the 1.4~GHz emitting region is well above $R$ = 124 light days.
This large size can well explain the lack of significant variability on time scales of tens of days.
We did not detect significant variability in our radio monitoring covering a time scale of $\sim$ 300 days, which may imply that the optically thin source is even larger than $\sim$ 300 light days or does not vary during this period.

The X-ray light curves also exhibit different behaviours in these three objects.
Mrk~110 generally shows little variability on time scales of days, and significant variability of increasing amplitudes with increasing time scales, from tens of days to hundreds of days.
A similar variability pattern is found in a daily monitoring campaign of Mrk~110 with \textit{Swift} XRT for a period of more than 200 days \citep{Vincentelli2021}.
Mrk~766 also varies significantly on time scales of days, but it also shows variability on longer time scales of tens of days.
An analysis of \textit{XMM-Newton} archival data shows that Mrk~766 exhibits fast variability on short time scales from hours to days, and large amplitude variability on long time scales of years \citep{Giacche2014}.
NGC~4593 shows fast variability on a time scale of a couple of days, and does not show significant variability on longer time scales.
A \textit{Swift} XRT monitoring campaign for a period of 23 days observed NGC~4593 every 1--2 orbits (96 min each orbit), and found that the object varies on short time scales from a couple of hours to a couple of days \citep{McHardy2018}.

\subsection{The radio versus X-ray delay}

\begin{figure*}
\centering
\includegraphics[width=.325\textwidth, trim={0.5cm 0cm 1cm 0cm}, clip]{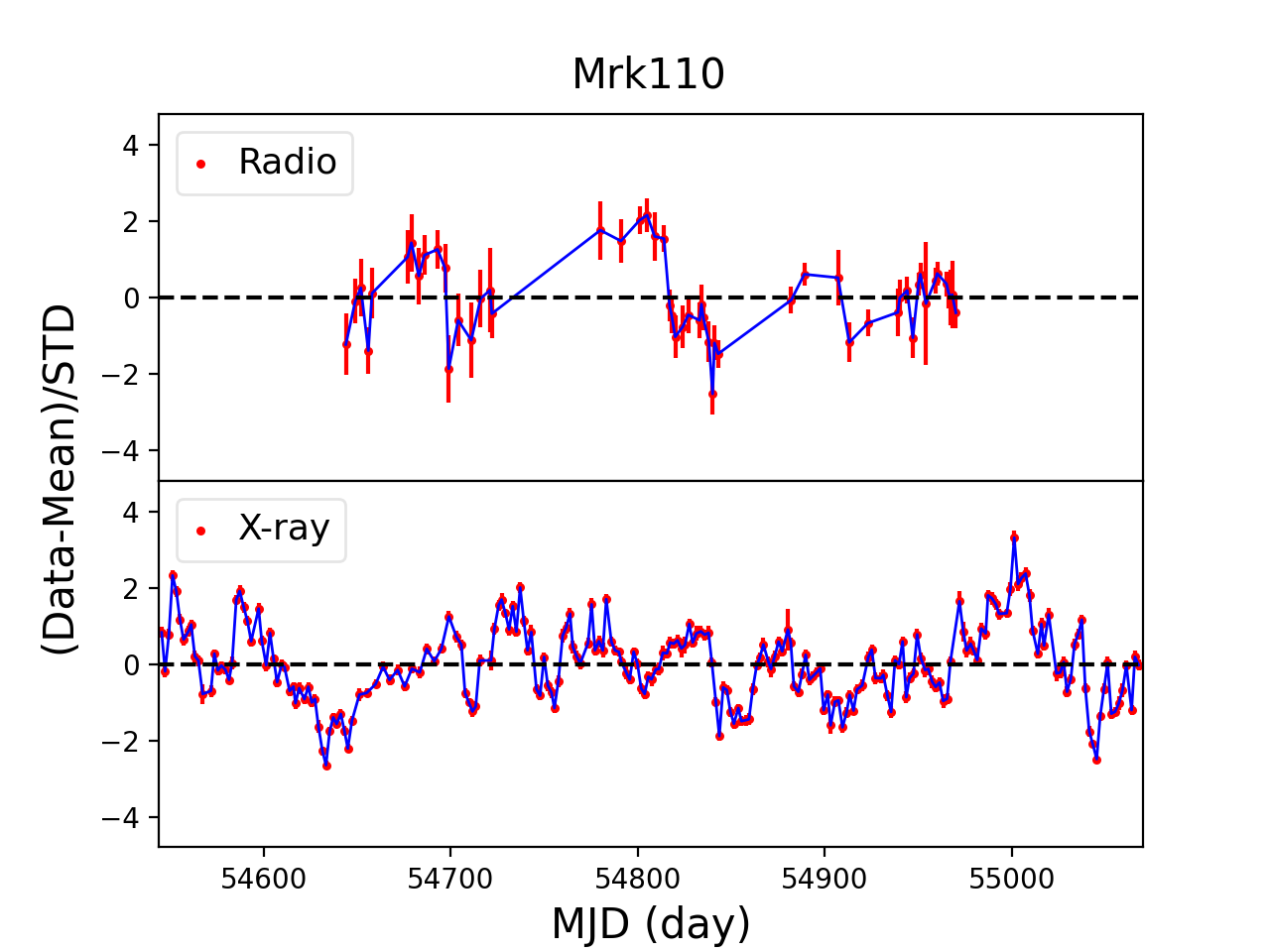}
\includegraphics[width=.325\textwidth, trim={0.5cm 0cm 1cm 0cm}, clip]{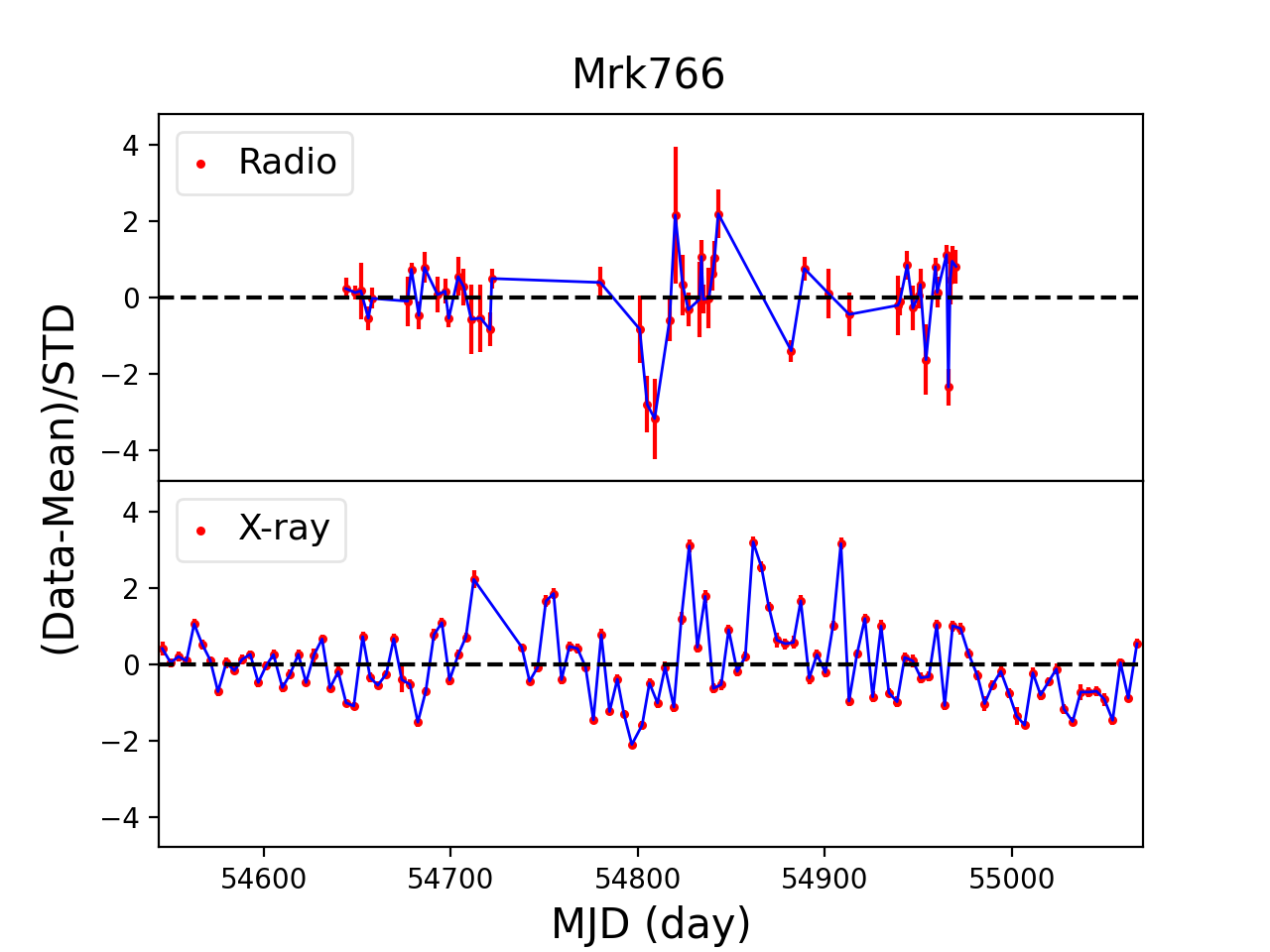}
\includegraphics[width=.325\textwidth, trim={0.5cm 0cm 1cm 0cm}, clip]{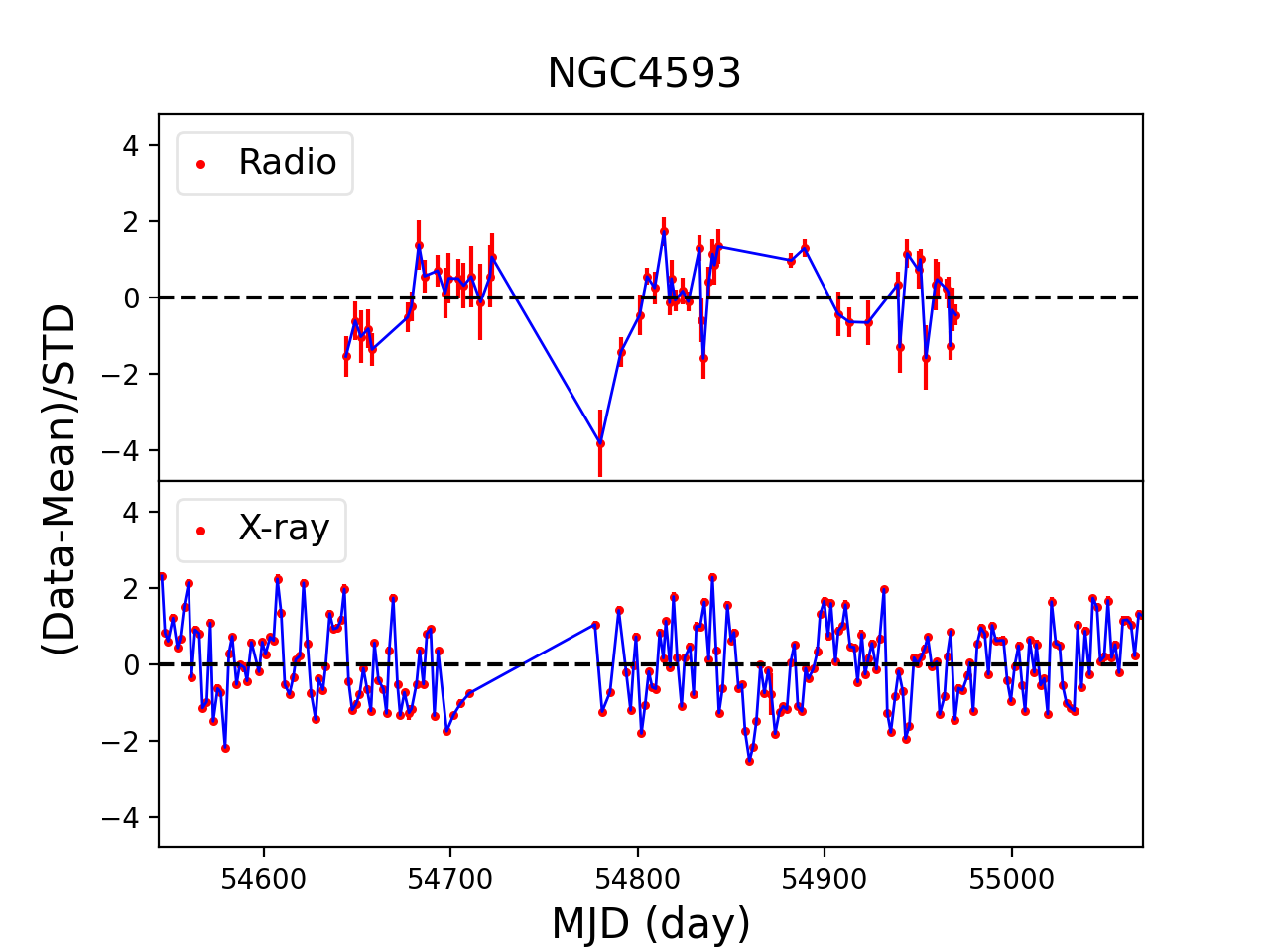}
\caption{The normalized radio peak flux density at 8.5~GHz (upper panel) and X-ray at 2--10~keV (lower panel) light curves of Mrk~110 (left), Mrk~766 (middle), and NGC~4593 (right).
The errors in the X-ray light curves are as small as the symbol size.}
\label{LC_RX}
\end{figure*}

In order to examine whether there is a time delay between the radio and X-ray light curves, we perform a Pearson cross-correlation analysis of the two light curves, as a function of the time delay.
Although other similar methods, such as ICCF, ZDCF, and JAVELIN \citep[see discussion and references in][]{Fian2022}, are commonly used to measure the delay time, and in particular the delay time uncertainty, our main aim here is to test the significance of the correlation, which these methods do not address.
To perform the analysis, we first subtract the mean flux density of $S_{\rm p}$ for the light curves in each configuration separately, then combine the light curves in the three configurations to form the full radio light curve of each object.
This combined radio light curve is then divided by its STD, to obtain a normalized radio light curve for each object.
The X-ray light curve, taken from $t_1 - 100$ to $t_2 + 100$ days, is separated into three periods, which engulf the periods of the three VLA configurations, and is then re-scaled as the radio light curve, that is by subtracting the mean flux of each period, and scaling the flux by the overall STD.
Figure~\ref{LC_RX} presents the normalized radio and X-ray light curves, which are used in the following analysis.

The value of Pearson cross-correlation coefficients $r$ is calculated for time lags from $t_1 - 100$ to $t_2 + 100$ days, with a time step of one day.
Since the radio light curves have a lower sampling rate than the X-ray light curves, we calculate the correlations using the observed radio fluxes, and the corresponding linearly interpolated X-ray fluxes.
We use the radio $S_{\rm p}$ light curve in the A configuration for Mrk~110 and NGC~4593, and the B configuration for Mrk~766, as these light curves show the most significant variability (smallest $p$ values).
The errors in the time lags are calculated via the Monte Carlo method.
This approach involves varying the radio flux density with a Gaussian-distributed noise proportional to the error in each epoch, and using the new radio light curve to cross-correlate with the X-ray one as previously mentioned.
The same process is repeated 1000 times, to estimate their STD.
In this way, we obtained a 1$\sigma$ error of the time lag.

\begin{figure}
\centering
\includegraphics[width=.5\textwidth]{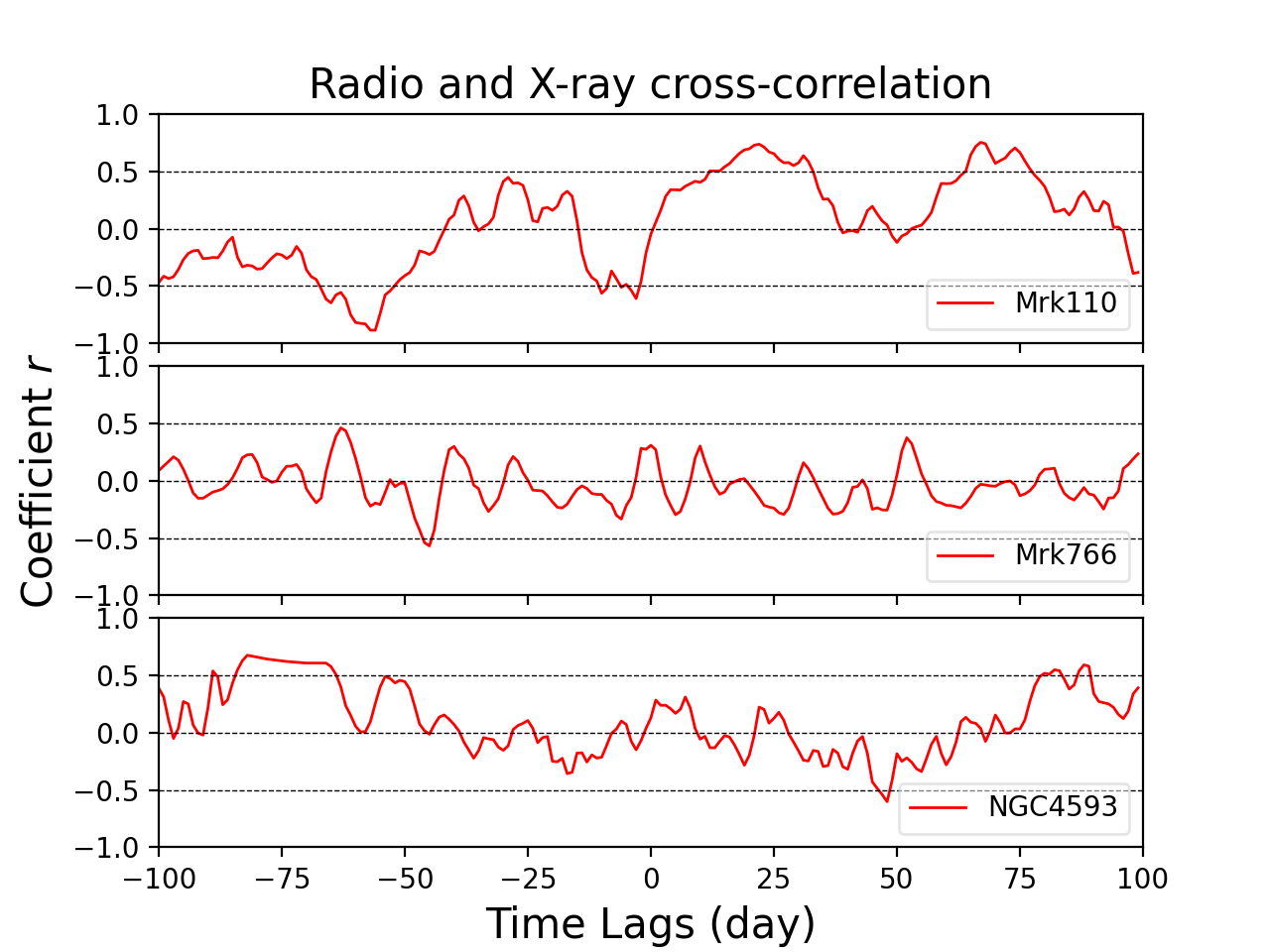}
\caption{The coefficients $r$ of the Pearson cross-correlation between the normalized radio and X-ray light curves for Mrk~110 (top panel), Mrk~766 (middle panel), and NGC~4593 (bottom panel). The radio peak flux densities in the A configuration for Mrk~110 and NGC~4593, and the B configuration for Mrk~766 are used.}
\label{r-value}
\end{figure}

Figure~\ref{r-value} presents the value of $r$ for Mrk~110, Mrk~766, and NGC~4593, as a function of the time delay, where a positive time lag means that the X-ray light curve lags the radio one.
Table~\ref{lags} lists the most apparently significant time lags, including both positive and negative correlations, and both positive and negative time delays.
The strongest correlations are found for Mrk~110, in particular a remarkably strong ($p = 10^{-6}$) negative ($r = -0.89$) correlation, where the radio lags the X-ray by 56 days.
The negative correlation between the radio and X-ray emission is opposite of the Neupert effect, but is not unexpected.
Since a coronal mass ejection event, if happens, may enhance the radio emission on large scales, but such an event depletes the coronal gas reservoir and thus reduces the coronal X-ray emission \citep{Jin2022}.

There are additional apparently significant time lags ($p = (4.7, 3) \times 10^{-4}$) in Mrk~110, with positive ($r = 0.74, 0.75$) correlations and positive delays where the X-ray lags the radio by 22 or 67 days.
In NGC~4593 there are weaker correlations ($r = -0.60, 0.59$) at a lower significance level ($p = (8.2, 9.5) \times 10^{-3}$), with positive delays of 48 or 88 days.
In Mrk~766 the strongest correlation is even weaker ($r = -0.57, p = 0.014$).

\begin{figure*}
\centering
\includegraphics[width=\textwidth, trim={3.5cm 1cm 3.5cm 1cm}, clip]{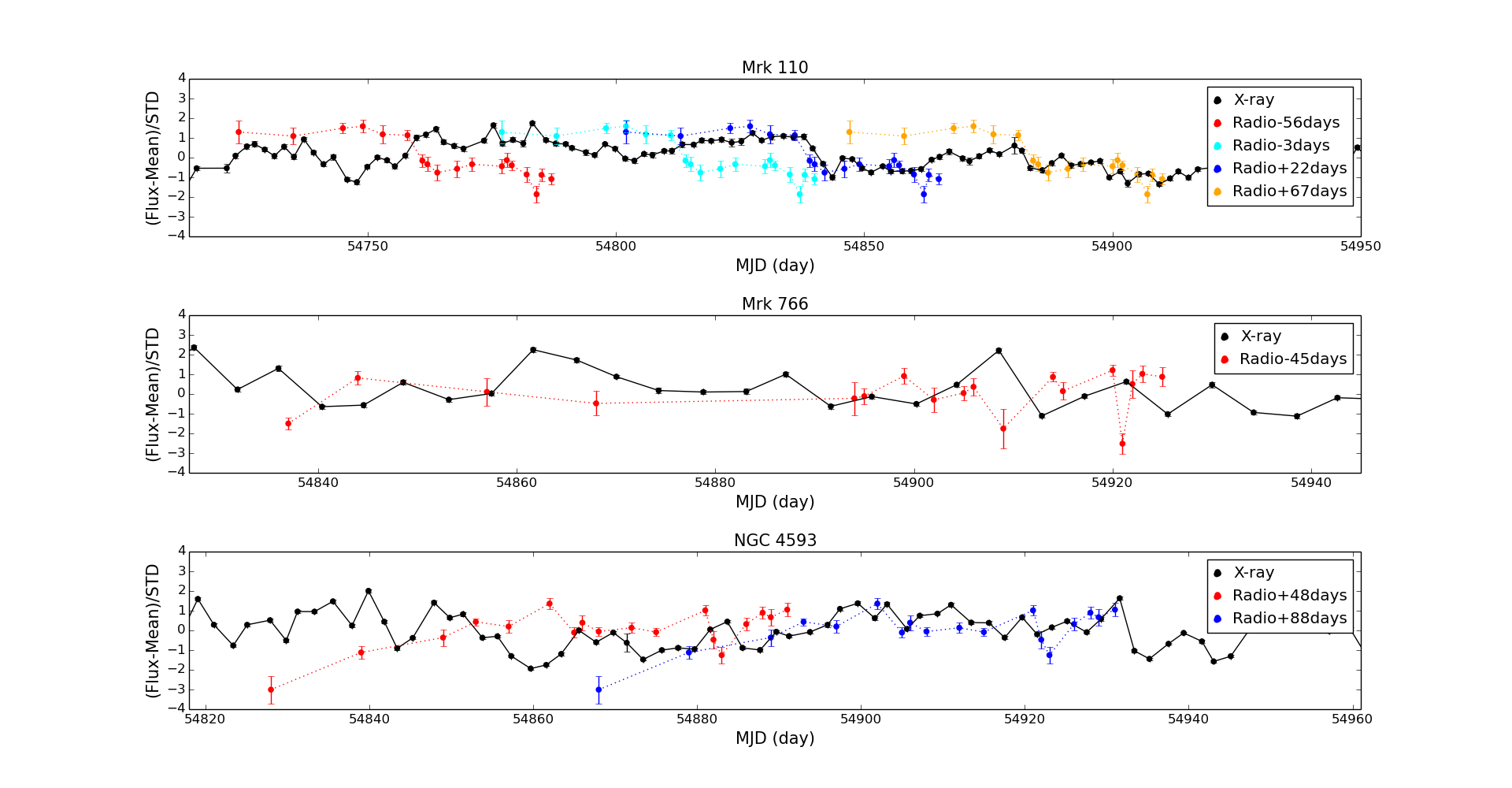}
\caption{The radio light curves shifted by the time lags in Table~\ref{lags} for Mrk~110 (top panel), Mrk~766 (middle panel), and NGC~4593 (bottom panel), relative to the peak flux densities in the A configuration for Mrk~110 and NGC~4593, and the B configuration for Mrk~766.
The errors in the X-ray light curves are as small as the symbol size.}
\label{moved}
\end{figure*}

Figure~\ref{moved} presents the X-ray light curve for each object, with the radio light curves shifted by the values given in Table~\ref{lags}.
Although the correlations for Mrk~110 appear highly significant, the light curves raise the suspicion that the correlations may be an artefact, produced by the similar variability characteristics of both light curves.
Specifically, both the radio and the X-ray light curves show roughly a step function on time scales of tens of days, which can match for a range of possible delays.
Also, the distribution of $r$ in Figure~\ref{r-value} seems to show an excess of high $r$ values.
Below we describe a more quantitative analysis, which shows that the significance of the $r$ values found above is much lower than derived from the Pearson correlation.

\begin{table}
\centering
\caption{Tentative time lags and Pearson cross-correlation.}
\label{lags}
\begin{tabular}{cccc}
\hline
\hline
Target & Time lag (day) & $r$-value & $p$-value \\
(1) & (2) & (3) & (4) \\
\hline
\multirow{3}{*}{Mrk~110} & $-$56 $\pm$ 1 & $-$0.89 & 1.0 $\times 10^{-6}$ \\
& $-$3 $\pm$ 3 & $-$0.61 & 7.3 $\times 10^{-3}$ \\
& $+$22 $\pm$ 3 & 0.74 & 4.7 $\times 10^{-4}$ \\
& $+$67 $\pm$ 3 & 0.75 & 3.0 $\times 10^{-4}$ \\
\hline
Mrk~766 & $-$45 $\pm$ 4 & $-$0.57 & 1.4 $\times 10^{-2}$ \\
\hline
\multirow{2}{*}{NGC~4593} & $+$48 $\pm$ 1 & $-$0.60 & 8.2 $\times 10^{-3}$ \\
& $+$88 $\pm$ 4 & 0.59 & 9.5 $\times 10^{-3}$ \\
\hline
\end{tabular}
\flushleft{\textbf{Notes.} Columns: (1) name, (2) time lag, (3) coefficient $r$, (4) probability $p$.
Positive time lags mean that the X-ray light curve lags the radio one, and vice versa.}
\end{table}

\subsection{A robustness test for $p(r)$}

As noted above, Figure~\ref{r-value} suggests an "over-abundance" of significant radio and X-ray delays for Mrk~110 within the range of delay times from $t_1 - 100$ to $t_2 + 100$ days.
Since the X-ray monitoring of Mrk~110 was carried out from 2005 to 2012, the X-ray light curve extends over more than 2000 days.
Both the theoretical and the observational radio variability time scales suggest that the emitting region extends over a region of the order of $\sim$ 10 light days across, while the more rapid X-ray variability suggests a significantly smaller region, $\sim$ 1 light day across or smaller.
It is therefore plausible to assume that a real delay of the variability in both bands is well below $\sim$ 100 light days.

The large time extent of the X-ray monitoring allows us to explore the distribution of the measured cross-correlation $p(r)$, that is the probability $p$ to find a correlation coefficient $r$ by chance over time scales of 2000 days, where we do not expect a real physical correlation.
Since the two light curves are not correlated, we expect the measured distribution $p(r)$ to follow the Pearson distribution for uncorrelated variables.

We therefore repeat the cross-correlation analysis by shifting the radio light curve from 1000 days before the first radio epoch ($t_1 - 1000$) to 1000 days following the last radio epoch ($t_2 + 1000$), with a time step of one day.
The $r$ value is calculated for the correlation in each time step between the normalized radio and X-ray light curves, excluding the 200 time steps in the range between $t_1 - 100$ and $t_2 + 100$ days, which leads to 1800 $r$ values for the uncorrelated data sets.

In order to increase the statistics, we reverse the time series of the radio light curve, and repeat the calculations of $r$.
This non-physical radio light curve reversed in time space, clearly cannot be physically correlated with the X-ray light curve.
This procedure adds 2000 more $r$ values to the uncorrelated data sets, which then gives a total of 3800 $r$ values.
The same procedures are applied for Mrk~766 and NGC~4593, where the available lengths of the X-ray light curves produce 3600 and 3800 $r$ values, respectively.

We order the distribution of $r$ values in each object with an increasing absolute value, and then calculate the measured $p(r)$, which equals the number of delays yielding a correlation $r$ larger than $r'$ with an absolute value ($|r| \geq |r'|$), divided by the total number of delays (3800 or 3600).
For each object we also calculate the Pearson distribution $p(r)$, which applies for the uncorrelated data sets.

\begin{figure*}
\centering
\includegraphics[width=.325\textwidth, trim={0.5cm 0cm 1cm 0cm}, clip]{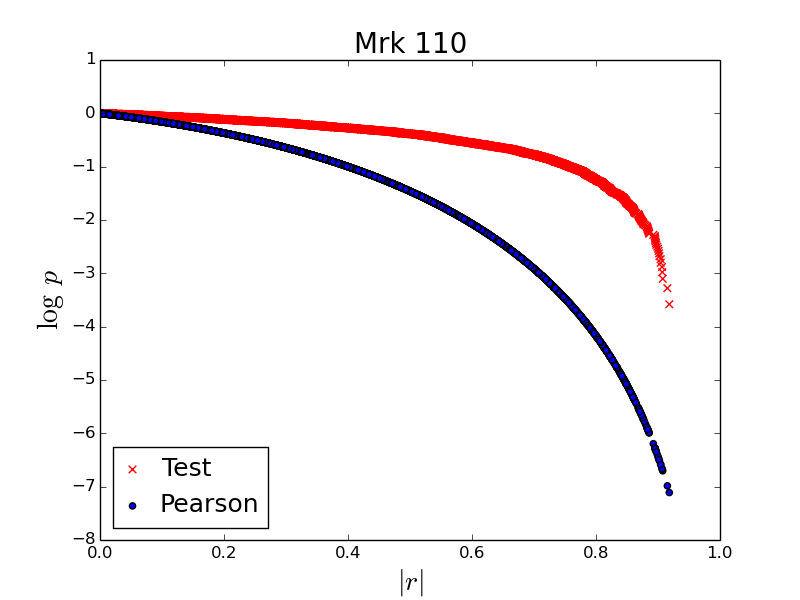}
\includegraphics[width=.325\textwidth, trim={0.5cm 0cm 1cm 0cm}, clip]{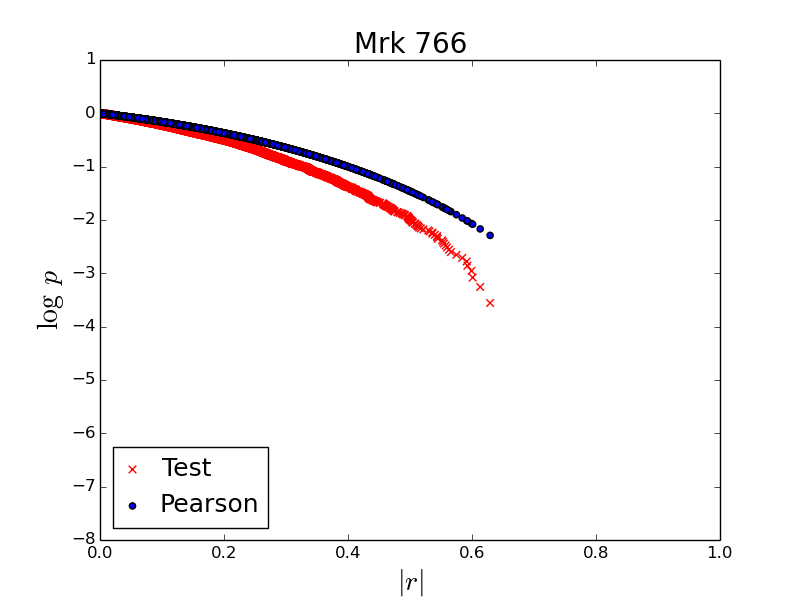}
\includegraphics[width=.325\textwidth, trim={0.5cm 0cm 1cm 0cm}, clip]{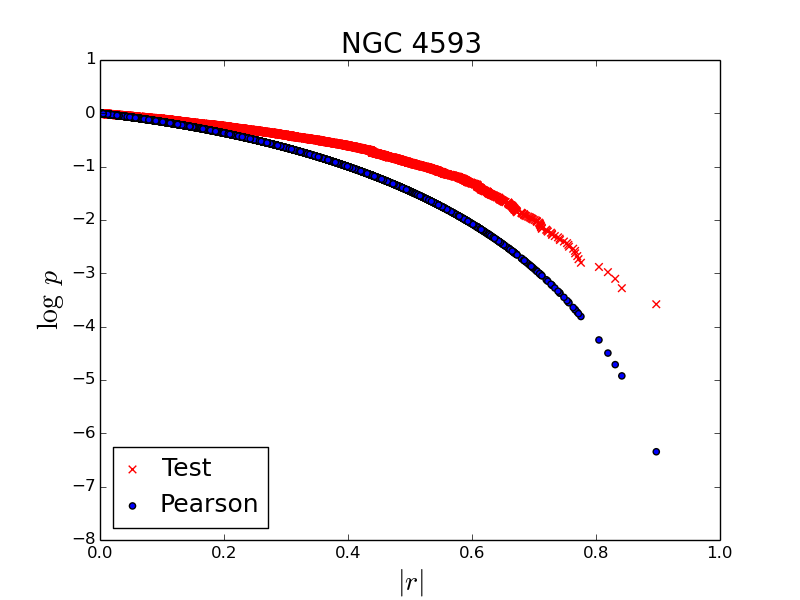}
\caption{The probability $\log p$ as a function of the coefficient $|r|$ of Mrk~110 (left panel), Mrk~766 (middle panel), and NGC~4593 (right panel) calculated from the Pearson cross-correlation and directly from our uncorrelated data sets.}
\label{test}
\end{figure*}

Figure~\ref{test} presents the directly measured $p(r)$ distribution and the expected Pearson $p(r)$ distribution for each object.
In Mrk~110, the measured $p(r)$ lies well above the Pearson $p(r)$ for all $r$ values, reaching a factor of $\sim 10^3$ for the highest $|r|$ values.
For example, the highest correlation of $r = -0.89$ found above, is formally highly significant with a Pearson $p = 1.0 \times 10^{-6}$.
However, there are 20 delays with $|r| \geq 0.89$ out of the 3800 delays tested, so the probability to find such a high correlation by chance in the uncorrelated data sets is actually 20/3800 = 0.5\%, which is a factor of $5 \times 10^3$ higher than the value from the Pearson correlation.
The probability of getting $|r| \geq 0.89$ by chance in the 200 correlations we made in the range between $t_1 - 100$ and $t_2 + 100$ days is therefore $\sim 0.5\% \times 200 = 100\%$.
Hence, the apparently highly significant delay in Mrk~110 with a remarkably high $r$ value, is actually completely insignificant.
Of course, all the other possible delays in this object are similarly insignificant.

To achieve a real correlation within the 200-day delays at a significant level of $>$ 95\%, the measured $p(r)$ needs to satisfy $p(r) \times 200 < 0.05$, that is $p(r) < 2.5 \times 10^{-4}$. This value cannot be reached in the simulations, which is limited by the total number of delays $\sim$ 4000 tested here.
Specifically, if a measured correlation has a $r$ value with a chance probability below 5\%, then it will not be found at all the simulations (4000) which is 20 times longer than the acceptable range of delays (200).
The simulations in fact reveal 20 cases with a higher $r$ values than found within the 200-day period, which is the expected number if the 200-day period are just a random subset of the $\sim$ 4000 days tested.

Why does the Pearson cross-correlation not yield the true distribution of $p(r)$ for the uncorrelated data sets in Mrk~110?
The Pearson distribution $p(r)$ applies for data sets of random numbers.
A time series of random numbers corresponds to a flat power spectrum, the so-called "white noise" spectrum.
The X-ray variability of AGN is often characterized by a steep power spectrum, the so-called "red noise" spectrum, where the variability amplitude increases on longer time scales.
This variability pattern is clearly present in the radio light curve of Mrk~110.
As a result, the fluxes at different time scales in a given light curve are correlated, in contrast with the underlying assumption of the Pearson correlation test, and this leads to a different $p(r)$ distribution when correlating two data sets with internal correlations.
For example, in the limit of a very steep power spectrum, the variability will be dominated by the lowest frequency which can be detected (set by the monitoring length), and can therefore be approximated by a $\delta$ function, that is a single $\sin$ wave.
So both the X-ray and the radio light curves will appear in such a case as a single $\sin$ wave, each one with a random phase since the two light curves are not physically related.
However, there will be a certain delay when the two light curves are in phase, yielding a high $|r|$ value.
The conclusion is that when cross-correlating "red noise" spectra, one cannot apply the Pearson distribution $p(r)$ to test the significance of the correlation.
This effect is properly acknowledged in some earlier studies \citep[e.g.][]{Uttley2003,Arevalo2008,Bell2011}, and is usually addressed by various simulations to derive the true $p(r)$.
Here, the great advantage is that the very long X-ray light curve allows us to measure directly the cross-correlation probability distribution of uncorrelated light curves, to derive the proper $p(r)$ distribution, and to test the significance of the $r$ values found in a more restricted range of possible delays.

In NGC~4593, the measured $p(r)$ of getting $|r| \geq 0.60$ is 183/3800 = 4.8\%, which is higher than the Pearson $p(r)$ of 0.8\%.
The variability amplitude in the radio light curve appears to slightly increase with time, that is a power spectrum slightly redder than "white noise", which may lead to a somewhat higher measured $p(r)$ compared to the Pearson $p(r)$.
In Mrk~766, the observed radio variability looks like "white noise", consistent with the small variability likely caused by the calibration uncertainty. The measured $p(r)$ is 8/3600 = 0.2\% to get $|r| \geq 0.57$, and the Pearson $p(r)$ is 1.4\%.
Though the measured $p(r)$ lies below the Pearson $p(r)$, which is opposite to the other two objects, given that the source is not variable, the mismatch is likely within the scatter expected for a specific realization.
Overall, the measured $p(r)$ in Mrk~766 is the closest one to the Pearson distribution.
The delays in both objects, with a Pearson $p(r)$ at a level of only $\sim 10^{-2}$, are thus completely insignificant given the 200-day delays tested in each object.

The conclusion is therefore that none of the present radio versus X-ray delays are significant.
More extended radio monitoring is required to reliably detect a temporal correlation with X-ray.

\section{Summary}

In this work, we present the results of quasi-simultaneous radio and X-ray monitoring of three RQ Seyfert galaxies, Mrk~110, Mrk~766, and NGC~4593. The radio observations were carried out with the VLA at 8.5~GHz and the X-ray observations were carried out with the RXTE at 2--10~keV. The results are summarized below.

1. After a detailed analysis of the integrated, peak, and extended flux densities of the radio light curves, we found that Mrk~110 observed with the A configuration, Mrk~766 observed with the B configuration, and NGC~4593 observed with the A configuration exhibited apparently significant radio variability with amplitudes of a few percent.
However, considering the VLA flux calibration uncertainty and the variability of the check-source, Mrk~110 appears to be marginally variable, Mrk~766 is not variable, and NGC~4593 shows the highest variability of $\sim$ 10\%.
The radio variability implies that the 8.5~GHz unresolved emission likely originates from a very compact core with a size of $\sim$ 10 light days, as expected for an optically thick source.

2. We performed a cross-correlation analysis of the radio peak flux density light curves, which show the most significant variability, with the X-ray light curves.
We found several tentative time lags.

3. However, following a further robustness test for the correlation significance, we found that the measured $p(r)$ distribution can deviate from the Pearson $p(r)$ distribution by as much as $\sim 10^3$ for Mrk~110.
This deviation results from the "red noise" characteristic of both the radio and the X-ray light curves, in contrast with the "white noise" assumed in the Pearson correlation test for uncorrelated data sets.
As a result, non of the potential delays found in this study are significant.

Finally, we highlight that the detection of radio variability requires high-resolution observations, in order to isolate the core emission.
In addition, longer period and higher time resolution observations are necessary to search for a physical connection between the radio and the X-ray emission.
The significance of the derived correlations should be tested by deriving the $p(r)$ distribution using Monte Carlo simulated data sets, \citep[e.g.][]{Uttley2003}.

\section*{Acknowledgements}

We thank Larry Rudnick and Phil Uttley for their very helpful and constructive comments concerning the significance of the delays.
We thank the anonymous referee for suggestions leading to the improvement of this work.
A.L. acknowledges support by the Israel Science Foundation (grant no.1008/18).
E.B. acknowledges support by a Center of Excellence of the Israel Science Foundation (grant no.2752/19).
The National Radio Astronomy Observatory is a facility of the National Science Foundation operated under cooperative agreement by Associated Universities, Inc.
This work has made use of light curves provided by the University of California, San Diego Center for Astrophysics and Space Sciences, X-ray Group (R.E. Rothschild, A.G. Markowitz, E.S. Rivers, and B.A. McKim), obtained at \url{https://cass.ucsd.edu/~rxteagn/}.

\section*{Data availability}

The radio data underlying this article are available in the NRAO Science Data Archive at \url{https://archive.nrao.edu/archive/advquery.jsp}, and can be accessed with the project codes of VLA/08B-182, VLA/08C-123, and VLA/09A-149.
The X-ray data underlying this article are available in the RXTE AGN Timing \& Spectral Database at \url{https://cass.ucsd.edu/~rxteagn/}.

\bibliographystyle{mnras}
\bibliography{arXiv.bbl}

\label{lastpage}
\end{document}